\documentclass[12pt]{article}
\usepackage{amsmath}
\usepackage{amsthm}
\usepackage{graphicx,psfrag,epsf}
\usepackage{enumerate}
\usepackage{natbib}
\usepackage{url} 
\usepackage{amssymb}
\usepackage{tabularx}
\usepackage{multirow}

\DeclareMathOperator*{\argmin}{arg\,min}

\newtheorem{theorem}{Theorem}

\newcommand{\blind}{0}

\begin{document}

\def\spacingset#1{\renewcommand{\baselinestretch}%
{#1}\small\normalsize} \spacingset{1}


\if0\blind
{
  \title{\bf Composite Expectile Regression with Gene-environment Interaction}
  \author{Jinghang Lin\\
    Department of Biostatistics, Yale University\\
    and \\
    Yuan Huang \\
   Department of Biostatistics, Yale University\\
   and \\
   Shuangge Ma*\\
   Department of Biostatistics, Yale University}
  \maketitle
} \fi

\if1\blind
{
  \bigskip
  \bigskip
  \bigskip
  \begin{center}
    {\LARGE\bf Composite Expectile Regression with Gene-environment Interaction}
\end{center}
  \medskip
} \fi

\bigskip
\begin{abstract}
If error distribution has 
heteroscedasticity, it voliates the assumption of linear regression. Expectile regression is a powerful tool for estimating the conditional expectiles of a response
variable in this setting.  Since multiple levels of  expectile regression model
has been well studied, we propose composite expectile regression by combining different levels of expectile regression to improve the efficacy. In this paper, we study the sparse
composite expectile regression under high dimensional setting. It is realized by implementing a coordinate descent algorithm. We also prove its selection and estimation consistency. Simulations are conducted to demonstrate its performance, which is comparable to or better than the alternatives.  We apply the proposed method to analyze Lung adenocarcinoma(LUAD) real data set, investigating the G-E interaction.  
\end{abstract}

\noindent%
\vfill

\newpage
\spacingset{1.45} 
\section{Introduction}
Linear regression minimizes the squared error
with one assumption: the variance of the noise terms is constant over all observations. In the real world data, the magnitude of the noise is not constant and the
data are heteroskedastic. Heteroscedasticity often
exists due to heterogeneity in measurement units or accumulation of outlying
observations from numerous sources of variables. Heteroscedasticity exists in biological data. For example, by implementing the genomics experiments, tens of thousands
of genes are often analyzed simultaneously by microarrays and occasional outlying
measurements appearing in numerous experimental and data preprocessing steps
can accumulate to form heteroscedasticity in the data obtained.

When we have heteroskedasticity in the error distribution, expectile regression with an asymmetric least 
squares(ALS) is proposed as a solution to it by Newey and Powell(1987). 
The main idea in expectile regression is to assign different
squared error loss to the positive and negative residuals, respectively. By 
doing
so, one can explores a complete relationship between the 
conditional expectile of a response variable and a set of 
predictor variables. Similar work has been done in quantile 
regression(Zou and Yuan(2008)). By adapting this composite loss 
function, composite quantile regression(CQR) is much more 
efficient than the LS estimator under many heavy-tailed error. Zhao and Xiao(2014) show that if  
we combine information over multiple quantiles, an 
upper bound on the distance between the efficiency of the 
estimator and the Fisher information decreases as the number of 
quantiles increases.  Motivated by 
composite quantile regression and related works, we proposed the 
composite expectile regression(CER) by combining information over 
different expectiles via a mix of ALS loss functions. The idea of 
CER is straightfoward: if more expectiles are used, we have more 
distributional information and can obtain more efficient 
estimation.

In this paper,  we develop sparse CER model under high dimensional setting since
  G-E interaction analysis is included. When fitting a sparse CQR 
  model, it is natural to adopt some classical penalties, such as 
  minmax concave penalty(MCP) by Zhang(2010). Coordinate
descent, the most popular algorithm for solving the least squares lasso, is used to this optimization problem. 
We derive the consistency of CER and show its application in detecting heteroscedasticity under high dimensional setting. For G-E interaction analysis, we assume that interaction term can not be identified if the corresponding main G effect is not identified. We decompose the G-E interaction coefficients to respect the "main effects, interactions" hierarchy. Minimax concave penalty(MCP) is applied to penalize genetic coefficient and G-E interaction coefficient.

The structure of the paper is as follows: In Section 2, we briefly
introduced expectile regression(ER) and composite expectile 
regression(CER).  The selection and estimation consistency of CER 
is presented in Section 3. We give a coordinate descent algorithm 
and run simulations to demonstrate its performance in Section 4. 
As an application, we apply CER and alternatives into analyzing lung adenocarcinoma data in section 5. Section 6 gives the summary and discussion.
\label{sec:intro}

\label{sec:meth}
\section{Methods}
In this section, we briefly introduce expectile regression and composite expectile regression under G-E interaction scenarios.  
\subsection{Expectile regression}
Consider a dataset with $n$ iid subjects. For the $i$th subject, 
let $Y_i$ be the response of interest, and 
$\pmb{Z_{i\cdot}} = (Z_{i1},...,Z_{iq})$ and 
$\pmb{X_{i\cdot}} = (X_{i1},...,X_{ip})$ be the 
$q-$ and $p-$dimensional vectors of Environmental(E) and Genetic(G) measurements. We consider 
the scenario with a continuous outcome and a expectile regression 
model with the joint effects of all E and G effects and their 
interactions:
\begin{align}
   Y_i = b + &\sum_{k=1}^q Z_{ik}\alpha_{k} + \sum_{j=1}^{p}X_{ij}\beta_j + \sum_{k=1}^{q}\sum_{j=1}^{p}Z_{ik}X_{ij}\eta_{kj}+\epsilon_{i}
\end{align}
where $b$ is intercept, $\lbrace \alpha_{k} \rbrace_{k=1,..,q}$, $\lbrace \beta_{j} \rbrace_{j=1,..,p}$ and 
$\lbrace \eta_{kj} \rbrace_{k=1,..,q, j=1,..,p}$ are the regression coefficients for 
the main E, main G, and G-E interactions, 
respectively, and $\lbrace \epsilon_{i} \rbrace_{i = 1,..,n}$ are the random 
errors.

To get the hierarchical constraint between main effects and interactions, we conduct the decomposition of $\eta_{kj}$ as $\eta_{kj}=\beta_{j}\gamma _{kj}$.
\begin{align}
   Y_i &= b + \sum_{k=1}^q Z_{ik}\alpha_{k} + \sum_{j=1}^{p}X_{ij}\beta_j + \sum_{k=1}^{q}\sum_{j=1}^{p}Z_{ik}X_{ij}\beta_{j}\gamma _{kj}+\epsilon_{i}\\
    & =b + \pmb{Z_{i.}}\pmb{\alpha}+\pmb{X_{i.}}\pmb{\beta}+\sum_{k=1}^{q}\pmb{M_{i.}}^{(k)}(\pmb{\beta}\odot \pmb{\gamma}_{k}) + \epsilon_i\\
    & = f(\pmb{X_{i.}}, \pmb{Z_{i.}}) + \epsilon_i, 
\end{align}
where $\pmb{\alpha}=(\alpha_{1},..,\alpha_{q})^{T
}$, $\pmb{\beta}=(\beta_1,...,\beta_{p})^{T}$,
$\pmb{\gamma_{k}}=(\gamma_{k1},..,\gamma_{kp})^{T}$, $\pmb{M_{i.}}^{(k)}=(Z_{ik}X_{i1},...,Z_{ik}X_{ip})$. And $\odot$ is the component-wise product. The coefficients of main G effects and G-E interaction effects are $\pmb{\beta}$ and $\pmb{\beta}\odot \pmb{\gamma}_{k}$. By adopting the decomposition technique of $\eta_{kj}$, we could guarantee that G-E interactions will not be identified if the corresponding main G effects are not identified. In other words, if $\beta_j=0$, then $\eta_{kj} = 0$.

Let $\pmb{Y} = (Y_1, ..., Y_n)^{T}$ be $n\times 1$ vector, $\pmb{b} = (b,...,b)^T$ be $n \times 1$ vector,  $\pmb{Z}=(\pmb{Z}_{1.}^{T},...,\pmb{Z}_{n.}^{T})^{T}$ be $n \times q $ matrix, $\pmb{X} = (\pmb{X}_{1.}^{T},...,\pmb{X}_{n.}^{T})^{T}$ be $n \times p$ matrix, $\pmb{M}^{(k)}= ((\pmb{M}_{1.}^{(k)})^{T},...,
(\pmb{M}_{n.}^{(k)})^{T})^{T}$ be $n \times p$ matrix. 
We have the following matrix form:
\begin{align*}
    \pmb{Y}=\pmb{b}+\pmb{Z\alpha}+\pmb{X\beta} + \sum_{k=1}^{q}\pmb{M}^{(k)}(\pmb{\beta}\odot \pmb{\gamma}_{k}) + \pmb{\epsilon},
\end{align*}
where $\pmb{\epsilon} = (\epsilon_1,...,\epsilon_n)^T$ is the vector of random error.

Here we consider the asymmetric square loss function. To get the estimators, we minimize the following
empirical risk function
\begin{align}
 \mathcal{Q}_{n}(\tau,\pmb{\theta})&=\frac{1}{2n}\sum_{i=1}^{n} L_{\tau}(y_i,f(\pmb{X_{i.}}, \pmb{Z_{i.}})) + \sum_{j=1}^{p}\rho(|\beta_j|;\lambda_{1}, r)\\
 &+ \sum_{j=1}^{p}\sum_{k=1}^{q}\rho(|\gamma_{kj}|;\lambda_{2},r), 0 <\tau < 1,
\end{align}

where $\pmb{\theta}  =(\pmb{b}^{T},\pmb{\alpha}^{T}, \pmb{\beta}^{T}, \pmb{\gamma}_{1}^{T},..., \pmb{\gamma}_{q}^{T})^{T}$. $\rho(|v|;\lambda, r) = \lambda\int_{0}^{|v|} \left( 1 - \frac{x}{\lambda r} \right)_{+} dx$ is the minimax concave penalty(MCP), where $r > 1$ is regularization parameter. $\lambda_1,\lambda_2$ are  two tuning parameters. And
\begin{equation}
L_{\tau}(Y_{i}, f(\pmb{X_{i.}}, \pmb{Z_{i.}}))=\left\{
\begin{aligned}
&(1- \tau) (Y_{i} - f(\pmb{X_{i.}}, \pmb{Z_{i.}}))^2, & if \ Y_{i} < f(\pmb{X_{i.}}, \pmb{Z_{i.}}) \\
&\tau (Y_i - f(\pmb{X_{i.}}, \pmb{Z_{i.}})))^2, & if \ Y_i \geq f(\pmb{X_{i.}}, \pmb{Z_{i.}}).
\end{aligned}
\right.
\end{equation}
Expectile regression can be regarded as weighted linear regression, but expectile regression has only two possible weight $\tau$ and $1-\tau$.
\subsection{Composite expectile regression}
To combine the strength across 
multiple expectile regression models, we propose composite 
expectile regression(CER) inspired by composite quantile 
regression(CQR) by Zou and Yuan(2008). 
Denote $0<\tau_{1}<...<\tau_{L}<1$. Specifically, we use the 
equally spaced expectiles: $\tau_{l}= \frac{l}{L+1}, 1 \leq l \leq L$. L is a 
positive constant and can be taken as 9 or 19. By adopting this 
strategy, we could improve
the efficiency.  We minimize the following objective function:
\begin{align}
\overline{\mathcal{Q}}_{n}(\pmb{\theta}) = \frac{1}{2n}\sum_{i=1}^{n}\sum_{l=1}^{L} L_{\tau_l}(Y_i,f(\pmb{X_{i.}}, \pmb{Z_{i.}})) + \sum_{j=1}^{p}\rho(|\beta_j|;\lambda_{1}, r)+ \sum_{j=1}^{p}\sum_{k=1}^{q}\rho(|\gamma_{kj}|;\lambda_{2},r),
\end{align}
where $f(\pmb{X_{i.}}, \pmb{Z_{i.}}) = b_{l} + \pmb{Z_{i.}}\pmb{\alpha}+\pmb{X_{i.}}\pmb{\beta}+\sum_{k=1}^{q}\pmb{M_{i.}}^{(k)}(\pmb{\beta}\odot \pmb{\gamma}_{k})$,
\begin{equation}
L_{\tau_{l}}(Y_{i}, f(\pmb{X_{i.}}, \pmb{Z_{i.}}))=\left\{
\begin{aligned}
&(1- \tau_{l}) (Y_{i} - f(\pmb{X_{i.}}, \pmb{Z_{i.}}))^2, & if \ Y_{i} < f(\pmb{X_{i.}}, \pmb{Z_{i.}}) \\
&\tau_{l} (Y_i - f(\pmb{X_{i.}}, \pmb{Z_{i.}})))^2, & if \ Y_i \geq f(\pmb{X_{i.}}, \pmb{Z_{i.}}).
\end{aligned}
\right.
\end{equation}

 Note that the regression coefficients are the same across different
expectile regression models, but intercepts are varied across different expectile regression models. Let $\pmb{Y} = (Y_1, ..., Y_n)^{T}$ be $n\times 1$ vector, $\pmb{b}_{l} = (b_{l},...,b_{l})^T$ be $n \times 1$ vector, $\pmb{Z}=(\pmb{Z}_{1.}^{T},...,\pmb{Z}_{n.}^{T})^{T}$ be $n \times q $ matrix, $\pmb{X} = (\pmb{X}_{1.}^{T},...,\pmb{X}_{n.}^{T})^{T}$ be $n \times p$ matrix, $\pmb{M}^{(k)}= ((\pmb{M}_{1.}^{(k)})^{T},...,
(\pmb{M}_{n.}^{(k)})^{T})^{T}$ be $n \times p$ matrix. We get $\pmb{\theta}  =(\pmb{b}_{1}^{T},...,\pmb{b}_{L}^{T},\pmb{\alpha}^{T}, \pmb{\beta}^{T}, \pmb{\gamma}_{1}^{T},..., \pmb{\gamma}_{q}^{T})^{T}$ by minimizing the penalized objective function in a matrix form:
\begin{align}
\overline{\mathcal{Q}}_{n}(\pmb{\theta}) &= \frac{1}{2n}\sum_{l=1}^{L}\mid \mid\pmb{W}^{1/2}_{\tau_{l}}(\pmb{Y}- \pmb{b}_{l} - \pmb{Z\alpha}-\pmb{X\beta} - \sum_{k=1}^{q}\pmb{M}^{(k)}(\pmb{\beta}\odot \pmb{\gamma}_{k}))\mid \mid^{2}_{2}\\
&+ \sum_{j=1}^{p}\rho(|\beta_j|;\lambda_{1}, r)+ \sum_{j=1}^{p}\sum_{k=1}^{q}\rho(|\gamma_{kj}|;\lambda_{2},r),
\end{align}
where $\pmb{W}_{\tau_l}$ is $n \times n$ diagonal matrix with two possible elements $\tau_{l}, 1- \tau_{l}$. For each element $w_{i}$ of $\pmb{W}_{\tau_{l}}$, $w_{i} = \tau_{l}$ if $Y_i > b_l + \pmb{Z_{i.}}\pmb{\alpha}+\pmb{X_{i.}}\pmb{\beta}+\sum_{k=1}^{q}\pmb{M_{i.}}^{(k)}(\pmb{\beta}\odot \pmb{\gamma}_{k})$, otherwise $w_i = 1- \tau_{l}.$

If G-E interaction coefficients are not decomposed, we denote it as non-hierarchical CER.  

\section{Statistical analysis}
In the section, we explore the consistency of CER with interaction term. We consider this scenario: when 
the sample size increases, the number of G factor increases and 
the number of E factor is finite. Let 
$\pmb{\theta}^{0} = 
\left((\pmb{b}_{1}^{0})^{T},...,(\pmb{b}_{L}^{0})^{T},(\pmb{\alpha}^{0})^{T},(\pmb{\beta}^{0})^{T},(\pmb{\gamma}_{1}^{0})^{T},...,(\pmb{\gamma}_{q}^{0})^{T}\right)^{T}$
be true parameter values. All $\lbrace\alpha_k\rbrace_{k=1,...,q}$ are not subjected to penalized and are nonzero. With hierarchical structure, we are only interested in those $\lbrace\gamma_{kj}\rbrace_{k=1,...,q,j=1,...,p}$ whose corresponding $\lbrace\beta_{j}\rbrace_{j=1,...,p}$ are nonzero. Let $\mathcal{A}_{1}= \lbrace j: \beta_{j}^{0} 
\neq 0 \rbrace$ be nonzero parameter of G effect, $\mathcal{A}_{2}^{k} = \lbrace j: \gamma_{kj}^{0} \neq 0 
\textit{ and } \beta_{j}^{0} \neq 0 \rbrace$ be nonzero parameter of  $k-$interaction effect.  With hierarchical structure, in each $\mathcal{A}_{2}^{k}$, $\gamma_{kj}$ is zero if the corresponding $\beta_{j}$ is zero too. For some $k, \text{ if } j \in \mathcal{A}_{2}^{k}$, then we have $j \in \mathcal{A}_{1}$. Let $\mathcal{A}_{2} = 
\mathcal{A}_{2}^{1}\cup ... \cup \mathcal{A}_{2}^{q}$,  $\mathcal{A}=\mathcal{A}_{1} \cup \mathcal{A}_{2}$ and $s= |\mathcal{A}_{1}| +|\mathcal{A}_{2}^{1}| + \dots + |\mathcal{A}_{2}^{q}|$. $\pmb{\theta}^{0}_{\mathcal{A}} = \left((\pmb{b}_{1}^{0})^{T},...,(\pmb{b}_{L}^{0})^{T},(\pmb{\alpha}^{0})^{T},(\pmb{\beta}_{\mathcal{A}_{1}}^{0})^{T},(\pmb{\gamma}_{\mathcal{A}^{1}_{2}}^{0})^{T},...,(\pmb{\gamma}_{\mathcal{A}^{q}_{2}}^{0})^{T}\right)^{T}$ is true parameter indexed by $\mathcal{A}$.

Denote $\hat{\pmb{\theta}}_{\mathcal{A}} = \left((\hat{\pmb{b}}_{1})^{T},...,(\hat{\pmb{b}}_{L})^{T}, (\hat{\pmb{\alpha}})^{T}, (\hat{\pmb{\beta}}_{\mathcal{A}_{1}})^{T},(\hat{\pmb{\gamma}}_{\mathcal{A}_{2}^{1}})^{T}...,(\hat{\pmb{\gamma}}_{\mathcal{A}_{2}^{q}})^{T}\right)^{T}$ as the minimizer of 
\begin{align*}
\overline{\mathcal{Q}}_{n}(\pmb{\theta}_{\mathcal{A}}) &= \frac{1}{2n}\sum_{l=1}^{L}\mid \mid\pmb{W}^{1/2}_{\tau_{l}}(\pmb{Y}-\pmb{b}_{l} - \pmb{Z\alpha}-\pmb{X}_{\mathcal{A}_1}\pmb{\beta}_{\mathcal{A}_{1}} - \sum_{k=1}^{q}\pmb{M}^{(k)}_{\mathcal{A}_{2}^{k}}(\pmb{\beta}_{\mathcal{A}_{2}^{k}}\odot \pmb{\gamma}_{k,\mathcal{A}_{2}^{k}}))\mid \mid^{2}_{2}\\
&+ \sum_{j=1}^{p}\rho(|\beta_j|;\lambda_{1}, r)+ \sum_{j=1}^{p}\sum_{k=1}^{q}\rho(|\gamma_{kj}|;\lambda_{2},r),
\end{align*}
where $\pmb{X}_{\mathcal{A}_1}$, $\pmb{\beta}_{\mathcal{A}_1}$ denote the components of $\pmb{X}$, $\pmb{\beta}$ indexed by $\mathcal{A}_1$, and  $\pmb{M}^{(k)}_{\mathcal{A}_{2}^{k}}$, $\pmb{\beta}_{\mathcal{A}_{2}^{k}}$, $\pmb{\gamma}_{k,\mathcal{A}_{2}^{k}}$ denote the components of $\pmb{M}^{(k)},\pmb{\beta},\pmb{\gamma}$ indexed by $\mathcal{A}_{2}^{k}$ for each $k$.

Suppose that we have the following conditions:
\begin{enumerate}
    \item $\epsilon$ are i.i.d and sub-Gaussian with noise level 
    $\sigma$. That is, for any vector $\mathbf{v}$ with 
    $||\mathbf{v}||_{2} = 1$ and any constant $t > 0$, 
    $P(|\mathbf{v}^{T}\epsilon|\geq t) \leq 2 
    exp(-\frac{t^2}{2\sigma^2})$.
    \item Let $b_0 = \min \lbrace \lbrace |\beta_{j}^{0}|: j \in 
    \mathcal{A}_{1} \rbrace, \lbrace |\gamma_{kj}^{0}|: j \in 
    \mathcal{A}_{2}^{k}, k = 1,...,q \rbrace \rbrace$, we have  $b_0 > a (\lambda_1 \vee \lambda_2), a > 0$ and $\lambda_1 \wedge \lambda_2 \gg \sqrt{s/n}$. 
    \item We use $\lambda_{min}(\cdot)$ and $\lambda_{max}(\cdot)$ to 
    represent the smallest and largest eigenvalues of a symmetric 
    matrix, respectively. Then $$ \max_{\pmb{\mathbf{\theta}}_{\mathcal{A}} 
    \in \mathcal{N}_{0}}\max_{l \in \lbrace 1,..., 
    L\rbrace}\lambda_{max}\left(\frac{1}{n}\pmb{G}(\pmb{\beta}_{\mathcal{A}_{2}},\pmb{\gamma}_{\mathcal{A}_{1}})^{T}\pmb{G}(\pmb{\beta}_{\mathcal{A}_{2}},\pmb{\gamma}_{\mathcal{A}_{1}}) \right) \leq 
    s\overline{c},$$  $$  \min_{\pmb{\mathbf{\theta}}_{\mathcal{A}} 
    \in \mathcal{N}_{0}}\min_{l \in \lbrace 1,..., 
    L\rbrace}\lambda_{min}\left(\frac{1}{n}\pmb{G}(\pmb{\beta}_{\mathcal{A}_{2}},\pmb{\gamma}_{\mathcal{A}_{1}})^{T}\pmb{G}(\pmb{\beta}_{\mathcal{A}_{2}},\pmb{\gamma}_{\mathcal{A}_{1}}) + \frac{1}{n}\pmb{F(\theta_{\mathcal{A}})} \right) \geq 
    \underline{c}, $$
    where $\pmb{\gamma}_{\mathcal{A}_1} = \left(  
    \pmb{\gamma}_{1,\mathcal{A}_1}^{T},.., 
    \pmb{\gamma}_{q,\mathcal{A}_{1}}^{T}\right)^{T}$ with $\gamma_{kj} = 0$, if $j \in \mathcal{A}_{1}$ but $j \notin \mathcal{A}_{2}^{k}$,
    $$\pmb{G}(\pmb{\beta}_{\mathcal{A}_2}, \pmb{\gamma}_{\mathcal{A}_1}) = \pmb{W}^{1/2}_{\theta_{
    \mathcal{A}}}\left(\pmb{1}_{n \times 1},\pmb{Z}, \pmb{U}(\pmb{\gamma}_{\mathcal{A}_1}), \pmb{V}^{(1)}(\pmb{\beta}_{\mathcal{A}_{2}^{1}}),...,\pmb{V}^{(q)}(\pmb{\beta}_{\mathcal{A}_{2}^{q}})\right)_{n \times(q+s+1)},$$ 
    with $$\pmb{U}(\pmb{\gamma}_{\mathcal{A}_1}) = \pmb{X}_{\mathcal{A}_1} +
    \sum_{k=1}^{q}\pmb{M}_{\mathcal{A}_1}^{(k)}\odot(\pmb{1}_{n\times 1} 
    (\gamma_{k,\mathcal{A}_1})^T),\pmb{V}^{(k)}(\pmb{\beta}_{\mathcal{A}_{2}^{k}}) = 
    \pmb{M}_{\mathcal{A}_{2}^{k}}^{(k)}\odot \left(\pmb{1}_{n \times 
    1}(\pmb{\beta}_{\mathcal{A}_{2}^{k}})^{T} \right).$$ $\pmb{F}_{\tau_{l}}(\pmb{\theta}_{\mathcal{A}}) = (f_{jh}(\pmb{\theta}_{\mathcal{A}}))_{(q+s)\times(q+s)}$ with $f_{jh}(\pmb{\theta}_{\mathcal{A}}) = -\pmb{W}_{\theta_{\mathcal{A}}}(\pmb{M}_{\varsigma}^{(k)})^{T}(\pmb{Y} - \pmb{b}_{l} - \pmb{Z\alpha} -\pmb{X}_{\mathcal{A}_{1}}\pmb{\beta}_{\mathcal{A}_{1}}-\sum_{g=1}^{q}\pmb{M}_{\mathcal{A}_{2}^{q}}^{(g)}(\pmb{\beta}_{\mathcal{A}_{2}^{g}}\odot \pmb{\gamma}_{g,\mathcal{A}_{2}^{g}}))$ if both $j$ and $h$ correspond to the $\varsigma$th element of $\mathcal{A}_{2}^{k}$, and 0 otherwise. For each element $w_{i}$ of $\pmb{W}_{\theta_{\mathcal{A}}}$, $w_{i} = \tau_{l}$ if $i$-th element  $[\pmb{Y}-\pmb{b}_{l} - \pmb{Z}\pmb{\alpha} - 
    \pmb{X}_{\mathcal{A}_1}\pmb{\beta}_{\mathcal{A}_1} -
    \sum_{k=1}^{q}\pmb{M}^{(k)}_{\mathcal{A}_{2}^{k}}(\pmb{\beta}_{\mathcal{A}_2^{k}}\odot (\pmb{\gamma}_{k,\mathcal{A}_{2}^{k}})]_{i} > 0$, otherwise $w_i = 1 - \tau_{l}$. 
    And $\overline{c}$ and $\underline{c}$ are positive constants, $\mathcal{N}_{0} = \lbrace \pmb{\theta}_{\mathcal{A}}:||\pmb{\theta}_{\mathcal{A}} - \pmb{\theta}^{0}_{\mathcal{A}}||_{\infty } \leq \frac{b_0}{2}\rbrace$. 
    \item Suppose$||\pmb{U}(\pmb{\gamma}_{\mathcal{A}_{1}^{c}}^{0})^{T}
    \pmb{G}(\pmb{\beta}_{\mathcal{A}_{2}}^{0},\pmb{\gamma}_{\mathcal{A}_{1}}^{0})||_{2,\infty}
    = O(n)$, $||\pmb{V}^{(k)}\left( \pmb{\beta}_{(\Tilde{A}_{2}^{k})^{c}} 
    \right)^{T}\pmb{G}(\pmb{\beta}_{\mathcal{A}_{2}^{0}}, 
    \pmb{\gamma}_{\mathcal{A}_{1}}^{0})||_{2,\infty} = O(n)$, 
    $||\pmb{U}(\pmb{\gamma}_{j}^{0})^{T}\hat{\pmb{W}}_{\tau_l}||_2 = O(\sqrt{n})$, 
    $||\pmb{V}^{(k)}(\pmb{\beta}_{j}^{0})^{T}\hat{\pmb{W}}_{\tau_l}||_2=O(\sqrt{n}), j= 
    1,..,p$, where each element $w_i$ of $\hat{\pmb{W}}_{\tau_{l}}$, $w_{i} = \tau_{l}$ if $i $-th element $\left( \pmb{Y}- \hat{\pmb{b}}_{l} - \pmb{Z}\hat{\pmb{\alpha}} - \pmb{X}\hat{\pmb{\beta}}- \sum_{k=1}^{q}\pmb{M}^{(k)}(\hat{\pmb{\beta}}\bigodot \hat{\pmb{\gamma}}_{k})\right)_i > 0$, otherwise, $w_i = 1 - \tau_{l}$. For $\pmb{Q}$, $||\pmb{Q}||_{2,\infty} = max_{||v||_{2} = 
    1}||\pmb{Q}v||_{\infty}$, $\mathcal{A}_{1}^{c} = \lbrace j: \beta_{j}^{0} = 0 \rbrace$ and $( \Tilde{\mathcal{A}}_{2}^{k} )^{c} = \lbrace j: \gamma_{kj}^{0} = 0 \text{ and } \beta_{j}^{0} \neq 0 \rbrace$. $\max_{\theta_{\mathcal{A}} \in 
    \mathcal{N}_{0}}\max_{j}\lambda_{max}\left(\pmb{T}_{1}^{(j)}(\gamma
    _{j})\right) = O(n)$, where $T_{1}^{(j)}(\gamma_{j}) =
    \left( t_{fh}^{(j)}(\gamma_{j})\right)_{(q+s)\times(q+s)}$ 
    with $t_{fh}^{(j)}(\gamma_{j}) = \sum_{l=1}^{L}\left( \pmb{X}_{j} + 
    \sum_{g=1}^{q}\pmb{M}_{j}^{(q)}\gamma_{gj} 
    \right)^{T}\hat{\pmb{W}}_{\tau_l}\pmb{M}_{\zeta}^{(k)}$, if both $f$ and $h$ correspond 
    to the $\zeta$th element of $\mathcal{A}_{2}^{k}$, and 0 
    otherwise. $\pmb{T}_{2}^{(j)}(\beta_{j}) = \left( 
    t_{fh}^{(j)}(\beta_{j})\right)_{(q+s)\times(q+s)}$ with 
    $t_{fh}^{(j)}(\beta_{j}) = \left( \pmb{M}_{j}^{(k)}\beta_{j} 
    \right)^{T}\pmb{M}_{\zeta}^{(k)}$, if both $f$ and $h$ correspond 
    to the $\zeta$th element of $\mathcal{A}_{2}^{k}$, and 0 
    otherwise. 
    \item $log(p) = O(n^{a}), a \in (0,1/2).$
    \item $\frac{\lambda_{i}}{\sqrt{s/n}} \to \infty, 
    \frac{\lambda_{i}}{n^{a/2 -1/2}\sqrt{log n}} \to \infty, i= 1,2. $
    \item $\frac{b_0}{\lambda_1} \to \infty.$
\end{enumerate}
Condition 1 is commonly assumed in the literature of high dimensional statistics. See for example Fa and Lv(2010). Condition 2 supposes that a smallest signal of genetic coefficient and G-E interaction coefficient with a rate that is not faster than $\sqrt{n/s}$. Condition 3 assumes that the eigenvalue of  design matrix is bounded, away from zero and infinity. The form of condition 3 is more complicated because of the decomposition technique to satisfy the hierarchy situation. Without decomposition of $\eta_{kj}$, condition 3 will be simpler with $\pmb{F(\theta_{\mathcal{A}})} = 0$. Condition 4 is similar to Condition 6 in Wu, Zhang and Ma(2020). The first two equations assume a relationship between negligible variables(in $\mathcal{A}_{1}^{c}$ and $(\Tilde{\mathcal{A}}_{2}^{k})^{c}$) and significant variables(in $\mathcal{A}_{1}$ and $\mathcal{A}_{2}$). Condition 5 assumes that the number of genetic coefficient has nonpolynomial dimensionality. See Fan and Lv(2010) for an overview of Variable Selection in high dimensional feature space. Condition 6 restricts the order of tuning pararmeters $\lambda_1, \lambda_2$. Condition 7 restricts the nonzero coefficients away from zero(Huang et al., 2017).

Next, we want to establish estimation consistency when the true sparsity structures are known.
\begin{theorem}
Under condition 1-3, there exists a local minimizer $\hat{\pmb{\theta}}_{\mathcal{A}}$ of $\overline{\mathcal{Q}}_{n}(\pmb{\theta}_{\mathcal{A}})$ such that for any constant $C>0$,
\begin{equation}
  P( ||\hat{\pmb{\theta}}_{\mathcal{A}} - \pmb{\theta}_{\mathcal{A}}^{0}||_{2}\leq \delta_{n}) > 1 - \eta_1.
\end{equation}
where $\delta_n =C\sqrt{s/n}$, $\eta_1 = L\cdot exp\left(-\frac{C\sqrt{n/s}\underline{c}^2}{32\overline{c}\sigma^2}\right)$, $L$ is specified as a constant in loss function.
\end{theorem}
Proof: it is sufficient to prove 
$$
P\lbrace \inf_{\pmb{\theta}_{\mathcal{A}} \in \mathcal{N}_{1}} \overline{\mathcal{Q}}_{n}(\pmb{\theta}_{\mathcal{A}}) > \overline{\mathcal{Q}}_{n}(\pmb{\theta}_{\mathcal{A}}^{0}) \rbrace \geq 1 - \eta_1,
$$
where $\mathcal{N}_{1} = \lbrace \pmb{\theta}_{\mathcal{A}}: ||\pmb{\theta}_{\mathcal{A}} - \pmb{\theta}_{\mathcal{A}}^{0}||_{2}  = \delta_{n} \rbrace$.

Let $\pmb{\omega}= \left((\pmb{e}_{1_{n \times 1}})^{T},...,(\pmb{e}_{L_{n \times 1}})^{T},(\pmb{g}_{q\times1})^{T}, (\pmb{u}_{|\mathcal{A}_{1}|\times1})^{T}, (\pmb{v}_{1_{|\mathcal{A}_{2}^{1}|\times 1}})^{T},..,(\pmb{v}_{q_{|\mathcal{A}_{2}^{q}|\times 1}})^{T} \right)^{T}$ with $||\pmb{\omega}||_{2} = 1$ and $\pmb{\theta}_{\mathcal{A}} = \pmb{\theta}_{\mathcal{A}}^{0}+\delta_n\pmb{\omega}$. 

Suppose $L_{\tau_l}(\pmb{\theta}_{\mathcal{A}}) = ||\pmb{W}_{\tau_l}^{1/2}\left(\pmb{Y} - \pmb{b}_{l} - \pmb{Z}\pmb{\alpha} - \pmb{X}_{\mathcal{A}_{1}}\pmb{\beta}_{\mathcal{A}_{1}} - \sum_{k=1}^{q}\pmb{M}_{\mathcal{A}_{2}^{k}}^{(k)}(\pmb{\beta}_{\mathcal{A}_{2}^{k}}\odot\pmb{\gamma}_{k,\mathcal{A}_{2}^{k}}) \right)||^{2}_{2}$, then we have
\begin{align*}
D_{n}(\pmb{\omega}) &= 
\overline{\mathcal{Q}}_{n}(\pmb{\theta}_{\mathcal{A}}^{0} + \delta_{n}\pmb{\omega}) 
- \overline{\mathcal{Q}}_{n}(\pmb{\theta}_
{\mathcal{A}}^{0})\\
& = \left[\frac{1}{2n}\sum_{
l=1}^{L} L_{\tau_l}(\pmb{\theta}_{\mathcal
{A}}^{0} + \delta_{n}\pmb{\omega})  -  
\frac{1}{2n}\sum_{l=1}^{L} 
L_{\tau_l}(\pmb{\theta}_{\mathcal{A}}^{0})
\right]\\
& = \frac{1}{2n}\sum_{l=1}^{L}\delta_{n}\pmb{\omega}^{T}\left(\nabla L_{\tau_{l}}(\pmb{\theta}_{\mathcal{A}})|_{\pmb{\theta}_{\mathcal{A}}^{0}}\right) + \frac{1}{4n}\sum_{l=1}^{L}\delta_{n}^{2}\pmb{\omega}^{T}\left(\nabla^{2} L_{\tau_{l}}(\pmb{\theta}_{\mathcal{A}})|_{\Tilde{\pmb{\theta}}_{\mathcal{A}}}\right)\pmb{\omega}\\
&= \delta_n\pmb{\omega}^{T}\sum_{l=1}^{L}\left[ -\frac{1}{n}\pmb{G}_{\tau_{l}}(\pmb{\beta}_{\mathcal{A}_{2}}^{0}, \pmb{\gamma}_{\mathcal{A}_{1}}^{0})^{T}\pmb{\epsilon}_{l}\right] + \frac{1}{2}\delta_{n}^{2}\pmb{\omega}^{T}\sum_{l=1}^{L} \left[ \frac{1}{n}\pmb{G}_{\tau_l}(\Tilde{\pmb{\beta}}_{\mathcal{A}_{2}},\Tilde{\pmb{\gamma}}_{\mathcal{A}_{1}})^{T}\pmb{G}_{\tau_l}(\Tilde{\pmb{\beta}}_{\mathcal{A}_{2}},\Tilde{\pmb{\gamma}}_{\mathcal{A}_{1}}) + \frac{1}{n}\pmb{F}_{\tau_{l}}(\Tilde{\pmb{\theta}}_{\mathcal{A}}) \right]\pmb{\omega}\\
& = T_1 + T_{2},
\end{align*}
where $\pmb{\epsilon}_{l} =\pmb{W}^{1/2}_{\pmb{\theta}^{0}_{\mathcal{A}}}\left[ \pmb{Y}-\pmb{b}^{0}_{l} - \pmb{Z\alpha}^{0} - \pmb{X}_{\mathcal{A}_1}\pmb{\beta}^{0}_{\mathcal{A}_1} -\sum_{k=1}^{q} \pmb{M}^{(k)}_{\mathcal{A}_{2}^{k}}(\pmb{\beta}_{\mathcal{A}_{2}^{k}}^{0}\odot\pmb{\gamma}_{k,\mathcal{A}_{2}^{k}}^{0}) \right]$, 
    $\pmb{\gamma}_{\mathcal{A}_1} = \left(  
    \pmb{\gamma}_{1,\mathcal{A}_1}^{T},.., 
    \pmb{\gamma}_{q,\mathcal{A}_{1}}^{T}\right)^{T}$ with $\gamma_{kj} = 0$, if $j \in \mathcal{A}_{1}$ but $j \notin \mathcal{A}_{2}^{k}$,
    $$
    \pmb{G}_{\tau_{l}}(\pmb{\beta}_{\mathcal{A}_2}, \pmb{\gamma}_{\mathcal{A}_1}) = \pmb{W}^{1/2}_{\pmb{\theta}_{
    \mathcal{A}}}\left(\pmb{1}_{n \times 1}, \pmb{Z}, \pmb{U}(\pmb{\gamma}_{\mathcal{A}_1}), \pmb{V}^{(1)}(\pmb{\beta}_{\mathcal{A}_{2}^{1}}),...,\pmb{V}^{(q)}(\pmb{\beta}_{\mathcal{A}_{2}^{q}})\right)_{n \times(q+s+1)},
    $$ 
    and   
    $$
    \pmb{U}(\pmb{\gamma}_{\mathcal{A}_1}) = \pmb{X}_{\mathcal{A}_1} +
    \sum_{k=1}^{q}\pmb{M}_{\mathcal{A}_1}^{(k)}\odot(\pmb{1}_{n\times 1} 
    (\gamma_{k,\mathcal{A}_1})^T),\pmb{V}^{(k)}(\pmb{\beta}_{\mathcal{A}_{2}^{k}}) = 
    \pmb{M}_{\mathcal{A}_{2}^{k}}^{(k)}\odot \left(\pmb{1}_{n \times 
    1}(\pmb{\beta}_{\mathcal{A}_{2}^{k}})^{T} \right).
    $$ 
    $\pmb{F}_{\tau_{l}}(\pmb{\theta}_{\mathcal{A}}) = (f_{jh}(\pmb{\theta}_{\mathcal{A}}))_{(q+s)\times(q+s)}$ with $f_{jh}(\pmb{\theta}_{\mathcal{A}}) = -\pmb{W}_{\theta_{\mathcal{A}}}(\pmb{M}_{\varsigma}^{(k)})^{T}(\pmb{Y} - \pmb{b}_{l }- \pmb{Z\alpha} -\pmb{X}_{\mathcal{A}_{1}}\pmb{\beta}_{\mathcal{A}_{1}}-\sum_{g=1}^{q}\pmb{M}_{\mathcal{A}_{2}^{q}}^{(g)}(\pmb{\beta}_{\mathcal{A}_{2}^{g}}\odot \pmb{\gamma}_{g,\mathcal{A}_{2}^{g}}))$ if both $j$ and $h$ correspond to the $\varsigma$th element of $\mathcal{A}_{2}^{k}$, and 0 otherwise. And $\Tilde{\pmb{\theta}}_{\mathcal{A}}$ lies on the line segment connecting $\pmb{\theta}^{0}_{\mathcal{A}}+\delta_n\pmb{\omega}$ and $\pmb{\theta}^{0}_{\mathcal{A}}$. For each element $w_{i}$ of $\pmb{W}_{\theta_{\mathcal{A}}}$, $w_{i} = \tau_{l}$ if $i$-th element  $[\pmb{Y}-\pmb{b}_{l} - \pmb{Z}\pmb{\alpha} - 
    \pmb{X}_{\mathcal{A}_1}\pmb{\beta}_{\mathcal{A}_1} -
    \sum_{k=1}^{q}\pmb{M}^{(k)}_{\mathcal{A}_{2}^{k}}(\pmb{\beta}_{\mathcal{A}_2^{k}}\odot (\pmb{\gamma}_{k,\mathcal{A}_{2}^{k}})]_{i} > 0$, otherwise $w_i = 1 - \tau_{l}$. 

Let
$
    T_1= \sum_{l=1}^{L}T_{1,l}
    =\sum_{l=1}^{L}\delta_n\pmb{\omega}^{T}\left[ -\frac{1}{n}\pmb{G}_{\tau_{l}}(\pmb{\beta}_{\mathcal{A}_{2}}^{0}, \pmb{\gamma}_{\mathcal{A}_{1}}^{0})^{T}\pmb{\epsilon}_{l}\right].
$
For $\epsilon_1 > 0$, with condition 1 about tail distribution of subgaussion variable and condition 3, we have
\begin{align*}
    P(T_{1,l} < -\delta_n\epsilon_1) &= P(\delta_n\pmb{\omega}^{T}\left[ -\frac{1}{n}\pmb{G}_{\tau_{l}}(\pmb{\beta}_{\mathcal{A}_{2}}^{0}, \pmb{\gamma}_{\mathcal{A}_{1}}^{0})^{T}\pmb{\epsilon}_{l}\right] < -\delta_n \epsilon_1)\\ 
    & = P\left(\frac{\delta_n\pmb{\omega}^{T}\left[ -\frac{1}{n}\pmb{G}_{\tau_{l}}(\pmb{\beta}_{\mathcal{A}_{2}}^{0}, \pmb{\gamma}_{\mathcal{A}_{1}}^{0})^{T}\pmb{\epsilon}_{l}\right]}{||\delta_n\pmb{\omega}^{T}\left[ -\frac{1}{n}\pmb{G}_{\tau_{l}}(\pmb{\beta}_{\mathcal{A}_{2}}^{0}, \pmb{\gamma}_{\mathcal{A}_{1}}^{0})^{T}\right]||_{2}} < - \frac{\epsilon_1}{||\delta_n\pmb{\omega}^{T}\left[ -\frac{1}{n}\pmb{G}_{\tau_{l}}(\pmb{\beta}_{\mathcal{A}_{2}}^{0}, \pmb{\gamma}_{\mathcal{A}_{1}}^{0})^{T}\right]||_{2}} \right)\\
    & \leq exp(-\frac{n\epsilon_{1}^{2}}{2\overline{c}s\sigma^2}).
\end{align*}
Set $\epsilon_1 = \frac{1}{4}\underline{c}\sqrt{\delta_n}$. With the Bonferroni’s inequality, we have
$$
P(T_1 < -\frac{1}{4}\sum_{l=1}^{L}\underline{c}\delta_n^{2}) \leq \sum_{l=1}^{L}P(T_{1,l} < -\frac{1}{4}\underline{c}\delta_{n}^{2}) \leq L exp\left(-\frac{n\underline{c}^2\delta_{n}}{32\overline{c}s\sigma^2} \right).
$$
Then 
$$
P(T_1 > -\frac{1}{4}L\underline{c}\delta_n^{2}) > 1-  L exp\left(-\frac{n\underline{c}^2\delta_{n}}{32\overline{c}s\sigma^2} \right).
$$
For $T_{2}$, with condition 2, we have
$$
||\Tilde{\pmb{\theta}}_{\mathcal{A}} - \pmb{\theta}_{\mathcal{A}}^{0}||_{\infty} \leq ||\pmb{\pmb{\theta}_{\mathcal{A}}} - \pmb{\theta}_{\mathcal{A}}^{0}||_{\infty} \leq \delta_{n} < b_{0}/2.
$$
By condition 3, we have
$$
T_{2}\geq \frac{1}{2}L\delta_{n}^2 \underline{c} >0.
$$

Let $\delta_n =C\sqrt{s/n}$, we have probability $1- L\cdot exp\left(-\frac{C\sqrt{n/s}\underline{c}^2}{32\overline{c}\sigma^2}\right)$ such that
 $D_n(\pmb{\omega}) \geq \frac{1}{4}L \delta^2_{n}\underline{c}.$ Then
\begin{align*}
P\lbrace \inf_{\pmb{\theta}_{\mathcal{A}} \in \mathcal{N}_{1}} \overline{\mathcal{Q}}_{n}(\pmb{\theta}_{\mathcal{A}}) > \overline{\mathcal{Q}}_{n}(\pmb{\theta}_{\mathcal{A}}^{0}) \rbrace&\geq P(D_{n}(\pmb{\omega})>0)\\
    & \geq P(\frac{1}{4}L\delta^2_{n}\underline{c}  \geq 0)\\
    & \geq P(T_{1} \geq \frac{1}{4}L\delta^2_{n}\underline{c})\\
    & \geq 1- L\cdot exp\left(-\frac{C\sqrt{n/s}\underline{c}^2}{32\overline{c}\sigma^2}\right).
\end{align*} 
 
Therefore, we prove the theorem 1. With Theorem 1, we have $||\hat{\pmb{\theta}}_{\mathcal{A}}-\pmb{\theta}_{\mathcal{A}}^{0}||_{2} = O_{p}(\sqrt{s/n})$. The order $O_{p}(\sqrt{s/n})$ is same to the  existing result by Wu, Zhang and Ma(2020). This theorem establishes estimation consistency of $\hat{\pmb{\theta}}_{\mathcal{A}}$.

Next, we establish the oracle  selection and estimation consistency properties of the proposed method. We check three condition in Theorem 1 in Fan and Lv (2011)[5]. 
Let $\mathcal{A}_{1}^{c} = \lbrace j: \beta_{j}^{0} = 0 \rbrace$ and $( \Tilde{\mathcal{A}}_{2}^{k} )^{c} = \lbrace j: \gamma_{kj}^{0} = 0 \text{ and } \beta_{j}^{0} \neq 0 \rbrace$. We have $( \Tilde{\mathcal{A}}_{2}^{k} )^{c} \cup \mathcal{A}_{1}^{c} = \lbrace j: \eta_{kj}^{0} = 0 \rbrace$.
Consider the oracle estimator $\hat{\pmb{\theta}}^{0}$ with $\hat{\pmb{\theta}}_{\mathcal{A}}^{0}= \hat{\pmb{\theta}}_{\mathcal{A}}$ and $\hat{\pmb{\theta}}_{\mathcal{A}^{c}}^{0}=0$. Theorem 2 provides sufficient conditions to ensure that $\hat{\pmb{\theta}}^{0}$ is a local minimizer of $\overline{\mathcal{Q}}_{n}(\pmb{\theta})$ with a high probability.
\begin{theorem}
Suppose $\hat{\pmb{\beta}}_{\mathcal{A}_{1}^{c}} = 0, \hat{\pmb{\gamma}}_{k,(\Tilde{\mathcal{A}}_{2}^{k})^{c}} =0$. Under condition 1-8, $\hat{\pmb{\theta}}$ is a strict local minimizer of $\overline{\mathcal{Q}}_{n}(\pmb{\theta})$ with probability approaching 1.
\end{theorem}
Proof: We follow Theorem 1 in Fan and Lv (2011)[5] to prove theorem 2. With Theorem 1 we have proven, it is sufficient to check condition (8) in the literature of Fan and Lv, which is equivalent to check Karush-Kuhn-Tucher(KKT) condition.

First, we consider $\hat{\beta}_{\mathcal{A}_{1}^{c}}$. Suppose
\begin{align*}
  h_1 = (n\lambda_1)^{-1}\left[\frac{1}{2}\sum_{l=1}^{L} \bigtriangledown_{\beta_{\mathcal{A}_{1}^{c}}}L_{\tau_l}(\pmb{\theta})\mid_{\hat{\pmb{\theta}}}\right]. 
\end{align*}

Since $\hat{\beta}_{\mathcal{A}_{1}^{c}} = 0$, by Taylor expansion, we have
\begin{align*}
    h_1 &= (n\lambda_1)^{-1}\left[\sum_{l=1}^{L}-\pmb{U}(\pmb{\gamma}_{\mathcal{A}_{1}^{c}})^{T}\hat{\pmb{W}}_{\tau_l}\left( \pmb{Y}- \hat{\pmb{b}}_{l}- \pmb{Z}\hat{\pmb{\alpha}} - \pmb{X}\hat{\pmb{\beta}}- \sum_{k=1}^{q}\pmb{M}^{(k)}(\hat{\pmb{\beta}}\bigodot \hat{\pmb{\gamma}}_{k})\right)\right]\\
    &= (n\lambda_1)^{-1}\left[\sum_{l=1}^{L} -\pmb{U}(\pmb{\gamma}_{\mathcal{A}_{1}^{c}}^{0})^{T}\hat{\pmb{W}}_{\tau_l}\pmb{\epsilon}_{l} +\left(\sum_{l=1}^{L}\pmb{U}(\pmb{\gamma}_{\mathcal{A}_{1}^{c}}^{0})^{T}\hat{\pmb{W}}_{\tau_l}\pmb{G}(\pmb{\beta}_{\mathcal{A}_{2}}^{0}, \pmb{\gamma}_{\mathcal{A}_{1}}^{0})^{T}(\hat{\pmb{\theta}}_{\mathcal{A}} - \pmb{\theta}_{\mathcal{A}}^{0})+\pmb{\kappa} \right)\right]\\
    & = (n\lambda_1)^{-1}\left[ I + II \right],
\end{align*}
where each element $w_i$ of $\hat{\pmb{W}}_{\tau_{l}}$, $w_{i} = \tau_{l}$ if $i $-th element $\left( \pmb{Y}- \hat{\pmb{b}}_{l} - \pmb{Z}\hat{\pmb{\alpha}} - \pmb{X}\hat{\pmb{\beta}}- \sum_{k=1}^{q}\pmb{M}^{(k)}(\hat{\pmb{\beta}}\bigodot \hat{\pmb{\gamma}}_{k})\right)_i > 0$, otherwise, $w_i = 1 - \tau_{l}$. And $\pmb{\epsilon}_{l} = \pmb{Y}- \pmb{b}^{0}_{l} - \pmb{Z\alpha}^{0} - \pmb{X}_{\mathcal{A}_1}\pmb{\beta}^{0}_{\mathcal{A}_1} -\sum_{k=1}^{q} \pmb{M}^{(k)}_{\mathcal{A}_{2}^{k}}(\pmb{\beta}_{\mathcal{A}_{2}^{k}}^{0}\odot\pmb{\gamma}_{k,\mathcal{A}_{2}^{k}}^{0}) $

 For II, let $m_{j}(\pmb{\theta}_{
 \mathcal{A}}) = \sum_{l=1}^{L}\left( \pmb{X}_j + \sum_{k=1}^{q}\pmb{M}_{j}^{(k)}\gamma_{kj} \right)^{T}\hat{\pmb{W}}_{\tau_l}\left(\pmb{b}_{l} + \pmb{Z}\pmb{\alpha} + \pmb{X}_{\mathcal{A}_{1}}\pmb{\beta}_{\mathcal{A}_{1}} + \sum_{k=1}^{q}\pmb{M}^{k}_{\mathcal{A}_{2}^{k}}(\pmb{\beta}_{\mathcal{A}_{2}^{k}}\odot\pmb{\gamma}_{k,\mathcal{A}_{2}^{k}})\right).$ Then $\pmb{\kappa} = \left(\kappa_j, j \in \mathcal{A}_{1}^{c} \right)^T$ with 
 \begin{align*}
     \kappa_j &= \frac{1}{2}(\hat{\pmb{\theta}}_{\mathcal{A}}- \pmb{\theta}_{\mathcal{A}}^{0})\left( \bigtriangledown_{\pmb{\theta}_{\mathcal{A}}}^{2}m_j(\pmb{\theta}_{\mathcal{A}})\mid_{\Tilde{\pmb{\theta}}_{\mathcal{A}}}\right)(\hat{\pmb{\theta}}_{\mathcal{A}}- \pmb{\theta}_{\mathcal{A}}^{0})\\
     & \leq \frac{1}{2}max_{j}\lambda_{max}(\pmb{T}_{1}^{(j)}(\Tilde{\pmb{\gamma}}_{j}))\vert \vert \hat{\pmb{\theta}}_{\mathcal{A}} - \pmb{\theta}_{\mathcal{A}}^{0} \vert \vert_{2},
 \end{align*}
 where $\Tilde{\pmb{\theta}}_{\mathcal{A}}$ lies on the line segment between $\hat{\pmb{\theta}}_{\mathcal{A}}$ and $\pmb{\theta}_{\mathcal{A}}^{0}$. And $\pmb{T}_{1}^{(j)}(\gamma_{j}) = \left( t_{fh}^{(j)}(\gamma_{j})\right)_{(q+s)\times(q+s)}$ with $t_{fh}^{j}(\gamma_{j}) = \sum_{l=1}^{L}\left( \pmb{X}_{j} + \sum_{g=1}^{q}\pmb{M}_{j}^{(q)}\gamma_{gj} \right)^{T}\hat{\pmb{W}}_{\tau_l}\pmb{M}_{\zeta}^{(k)}$, if both $f$ and $h$ correspond to the $\zeta$th element of $\mathcal{A}_{2}^{k}$, and 0 otherwise.
 With condition 4 and condition 6,
 \begin{align*}
     (n\lambda_1)^{-1}\vert \vert II \vert \vert_{\infty} &= (n\lambda_1)^{-1} \left(\vert \vert \sum_{l=1}^{L}\pmb{U}(\pmb{\gamma}_{\mathcal{A}_{1}^{c}}^{0})^{T}\hat{\pmb{W}}_{\tau_l}\pmb{G}(\pmb{\beta}_{\mathcal{A}_{2}}^{0}, \pmb{\gamma}_{\mathcal{A}_{1}}^{0})^{T}(\hat{\pmb{\theta}}_{\mathcal{A}} -\pmb{ \theta}_{\mathcal{A}}^{0})+\pmb{\kappa} \vert \vert_{\infty}\right)\\
     & = (n\lambda_1)^{-1} \left( O(n)\vert \vert \hat{\pmb{\theta}}_{\mathcal{A}} - \pmb{\theta}_{\mathcal{A}}^{0}  \vert \vert_{2} + O(n)\vert \vert \hat{\pmb{\theta}}_{\mathcal{A}} - \pmb{\theta}_{\mathcal{A}}^{0}  \vert \vert_{2}^{2} \right)\\
     &= O(\lambda_{1}^{-1}\sqrt{s/n}) = o(1).
 \end{align*}
 
 For I, consider the event
\begin{align*}
    \Omega_{1} = \lbrace \vert \vert  \pmb{U}(\gamma_{\mathcal{A}_{1}^{c}}^{0})^{T}\hat{\pmb{W}}_{\tau_l}\pmb{\epsilon}_{1} \vert \vert_{\infty} \leq \zeta_{n} \sqrt{n} \rbrace
\end{align*}
 with $\zeta_{n} = n^{a}(log(n))^{1/2}$. With Condition 4 and Condition 5, we have 
 \begin{align*}
     P(\Omega_{1}) &= 1- P\lbrace \vert \vert  \pmb{U}(\pmb{\gamma}_{\mathcal{A}_{1}^{c}}^{0})^{T}\hat{\pmb{W}}_{\tau_l}\pmb{\epsilon}_{1} \vert \vert_{\infty} > \zeta_{n} \sqrt{n} \rbrace \\
     & \geq 1- \sum_{j \in \mathcal{A}_{1}^{c}} P\lbrace \vert \vert  \pmb{U}(\pmb{\gamma}_{j}^{0})^{T}\hat{\pmb{W}}_{\tau_l}\pmb{\epsilon}_{1} \vert \vert_{\infty} > \zeta_{n} \sqrt{n} \rbrace\\
     & \geq 1 - 2p *exp \left(- \frac{\zeta_{n}^{2}n}{2\sigma^2\max_{j \in \mathcal{A}_{1}^{c}} \vert \vert \pmb{U}(\pmb{\gamma}_{j}^{0})^{T}\hat{\pmb{W}}_{\tau_l}\vert \vert^{2}_{2}} \right) \longrightarrow 1.
 \end{align*}
 as $log(p) = O(n^{a})$ and $\vert \vert \pmb{U}(\pmb{\gamma}_{j}^{0})^{T}\hat{\pmb{W}}_{\tau_l} \vert \vert_{2} = O(\sqrt{n})$. We have
 \begin{align*}
     \vert \vert \pmb{U}(\pmb{\gamma}_{\mathcal{A}_{1}^{c}}^{0})^{T}\hat{\pmb{W}}_{\tau_l}\pmb{\epsilon}_{1} \vert \vert_{\infty}  = O(n^{a/2 + 1/2}\sqrt{log n}).
 \end{align*}
 We use triangle inequality of maximal norm and L is a constant. Then
 \begin{align*}
     \vert \vert \sum_{l=1}^{L}\pmb{U}(\pmb{\gamma}_{\mathcal{A}_{1}^{c}}^{0})^{T}\hat{\pmb{W}}_{\tau_l}\pmb{\epsilon}_{1} \vert \vert_{\infty} \leq \sum_{l=1}^{L}\vert \vert \pmb{U}(\pmb{\gamma}_{\mathcal{A}_{1}^{c}}^{0})^{T}\hat{\pmb{W}}_{\tau_l}\pmb{\epsilon}_{1} \vert \vert_{\infty} = O(n^{a/2 + 1/2}\sqrt{log n}).
 \end{align*}
 With condition 6, we have
 \begin{align*}
     (n\lambda_{1})^{-1}\vert \vert \sum_{l=1}^{L}\pmb{U}(\pmb{\gamma}_{\mathcal{A}_{1}^{c}}^{0})^{T}\hat{\pmb{W}}_{\tau_l}\pmb{\epsilon}_{1} \vert \vert_{\infty} = o(1).
 \end{align*}

Next, we consider $\hat{\pmb{\gamma}}_{k, (\Tilde{{\mathcal{A}}}_{2}^{k})^{c}}.$ The steps are similar to $\hat{\pmb{\beta}}_{\mathcal{A}_{1}^{c}}$. Suppose
\begin{align*}
    h_{2} = (n\lambda_1)^{-1}\left[\frac{1}{2}\sum_{l=1}^{L} \bigtriangledown_{(\Tilde{\mathcal{A}}_{2}^{k})^{c}}L_{\tau_l}(\pmb{\theta})\mid_{\hat{\pmb{\theta}}} \right],
\end{align*}
Since $\hat{\pmb{\gamma}}_{k, (\Tilde{{\mathcal{A}}}_{2}^{k})^{c}} = 0$ and $\hat{\pmb{\beta}}_{(\Tilde{\mathcal{A}}_{2}^{k})^{c}} \neq 0,$ by Taylor expansion, we can get
\begin{align*}
    h_2 &= (n\lambda_2)^{-1}\left[\sum_{l=1}^{L}-\pmb{V}^{(k)}(\pmb{\beta}_{(\Tilde{\mathcal{A}}_{2}^{k})^{c}})^{T}\hat{\pmb{W}}_{\tau_l}\left( \pmb{Y}- \hat{\pmb{b}}_{l} - \pmb{Z}\hat{\pmb{\alpha}} - \pmb{X}\hat{\pmb{\beta}}- \sum_{k=1}^{q}\pmb{M}^{(k)}(\hat{\pmb{\beta}}\odot \hat{\pmb{\gamma}}_{k})\right)\right]\\
    &= (n\lambda_2)^{-1}\left[\sum_{l=1}^{L}-\pmb{V}^{(k)}(\beta_{(\Tilde{\mathcal{A}}_{2}^{k})^{c}})^{T}\hat{\pmb{W}}_{\tau_l}\epsilon + \left(\sum_{l=1}^{L}\pmb{V}^{(k)}(\pmb{\beta}_{(\Tilde{\mathcal{A}}_{2}^{k})^{c}})^{T}\hat{\pmb{W}}_{\tau_l}\pmb{G}(\pmb{\beta}_{\mathcal{A}_2}^{0},\pmb{\gamma}_{\mathcal{A}_{1}}^{0})^{T}(\hat{\pmb{\theta}}_{\mathcal{A}}-\pmb{\theta}_{\mathcal{A}}^{0})+\Tilde{\pmb{\kappa}}\right)\right]\\
    & = IV + V.
\end{align*}

For V, let $\Tilde{m}_{j}(\pmb{\theta}_{\mathcal{A}}) = \sum_{l=1}^{L}\left( \pmb{M}_{j}^{(k)}\beta_{j} \right)^{T}\hat{\pmb{W}}_{\tau_{l}}\left( \pmb{b}_{l} + \pmb{Z}\pmb{\alpha} + \pmb{X}_{\mathcal{A}_{1}}\pmb{\beta}_{\mathcal{A}_{1}} + \sum_{k=1}^{q}\pmb{M}_{\mathcal{A}_{2}^{(k)}}^{(k)}(\pmb{\beta}_{\mathcal{A}_{2}^{k}}\odot\pmb{\gamma}_{k,\mathcal{A}_{2}^{k}}) \right),$ then $\Tilde{\pmb{\kappa}} = (\Tilde{\kappa}_{j}, j \in (\Tilde{\mathcal{A}}_{2}^{k})^{c})^{T}$ with
\begin{align*}
    \Tilde{\kappa_j} &= \frac{1}{2}(\hat{\pmb{\theta}}_{\mathcal{A}}- \pmb{\theta}_{\mathcal{A}}^{0})\left( \bigtriangledown_{\pmb{\theta}_{\mathcal{A}}}^{2}m_j(\pmb{\theta}_{\mathcal{A}})\mid_{\Tilde{\pmb{\theta}}_{\mathcal{A}}}\right)(\hat{\pmb{\theta}}_{\mathcal{A}}- \pmb{\theta}_{\mathcal{A}}^{0})\\
     & \leq \frac{1}{2}max_{j}\lambda_{max}(\pmb{T}_{2}^{(j)}(\Tilde{\beta}_{j}))\vert \vert \hat{\pmb{\theta}}_{\mathcal{A}} - \pmb{\theta}_{\mathcal{A}}^{0} \vert \vert_{2},
\end{align*}
 where $\hat{\pmb{\theta}}_{\mathcal{A}}$ lies on the line segment between $\hat{\pmb{\theta}}_{\mathcal{A}}$ and $\pmb{\theta}_{\mathcal{A}}^{0}$, $\pmb{T}_{2}^{(j)}(\beta_{j}) = \left( t_{fh}^{(j)}(\beta_{j})\right)_{(q+s)\times(q+s)}$ with $t_{fh}^{j}(\beta_{j}) = \sum_{l=1}^{L}\left( \pmb{M}_{j}^{(k)}\beta_{j} \right)^{T}\hat{\pmb{W}}_{\tau_{l}}\pmb{M}_{\zeta}^{(k)}$, if both $f$ and $h$ correspond to the $\zeta$th element of $\mathcal{A}_{2}^{k}$, and 0 otherwise.
\begin{align*}
     (n\lambda_2)^{-1}\vert \vert V \vert \vert_{\infty} &= (n\lambda_2)^{-1} \left(\vert \vert \sum_{l=1}^{L}\pmb{V}^{(k)}(\pmb{\beta}_{\Tilde{\mathcal{A}}_{2}^{k}}^{0})^{T}\pmb{W}_{\tau_l}\pmb{G}(\pmb{\beta}_{\mathcal{A}_{2}}^{0}, \pmb{\gamma}_{\mathcal{A}_{1}}^{0})^{T}(\hat{\pmb{\theta}}_{\mathcal{A}} - \pmb{\theta}_{\mathcal{A}}^{0})+\pmb{\kappa} \vert \vert_{\infty}\right)\\
     & \leq (n\lambda_2)^{-1} \left( O(n)\vert \vert \hat{\pmb{\theta}}_{\mathcal{A}} - \pmb{\theta}_{\mathcal{A}}^{0}  \vert \vert_{2} + O(n)\vert \vert \hat{\pmb{\theta}}_{\mathcal{A}} - \pmb{\theta}_{\mathcal{A}}^{0}  \vert \vert_{2} \right)\\
     &\leq O(\lambda_{2}^{-1}\sqrt{s/n}) = o(1).
 \end{align*}
For IV, we consider the event
\begin{align*}
    \Omega_2 = \lbrace \vert \vert \pmb{V}^{(k)}(\pmb{\beta}_{(\Tilde{\mathcal{A}}_{2}^{k})^{c}})^{T}\hat{\pmb{W}}_{\tau_l}\pmb{\epsilon} \vert \vert_{\infty} \leq \zeta_{n} \sqrt{n} \rbrace,
\end{align*}
where $\zeta_n = n^{a}(log(n))^{1/2}$. Then,
\begin{align*}
    P(\Omega_2) &= 1 - P \lbrace \vert \vert \pmb{V}^{(k)}(\pmb{\beta}_{(\Tilde{\mathcal{A}}_{2}^{k})^{c}})^{T}\hat{\pmb{W}}_{\tau_l}\pmb{\epsilon} \vert \vert_{\infty} > \zeta_n \sqrt{n} \rbrace\\
    & \geq 1- \sum_{i \in (\Tilde{\mathcal{A}}_{2}^{k})^{c}} P \lbrace \vert \vert \pmb{V}^{(k)}(\pmb{\beta}_{i}^{0})^{T}\hat{\pmb{W}}_{\tau_l}\pmb{\epsilon} \vert \vert_{\infty} > \zeta_n \sqrt{n} \rbrace\\
    & \geq 1 - 2p* exp\left(-\frac{\zeta_{n}^{2}n}{2\sigma^2 max_{i \in (\Tilde{A}_{2}^{k})^{c}}\vert \vert \pmb{V}^{(k)}(\pmb{\beta}_{i}^{0})\hat{\pmb{W}}_{\tau_l}\vert \vert_{2}^{2}}\right) \longrightarrow 1.
\end{align*}
as $log(p) = O(n^a)$ and $\vert \vert \pmb{V}^{(k)}(\beta_{i}^{0})\hat{\pmb{W}}_{\tau_l}\vert \vert_{2} = O(\sqrt{n})$. Then we have, with probability approaching 1,
\begin{align*}
    \vert \vert \pmb{V}^{(k)}(\beta_{(\Tilde{\mathcal{A}}_{2}^{k})^{c}})^{T}\hat{\pmb{W}}_{\tau_l}\pmb{\epsilon} \vert \vert_{\infty} = O(n^{a/2+1/2}\sqrt{logn}).
\end{align*}
Since $L$ is a constant and triangle inequality of maximal norm, we have 
\begin{align*}
    \vert \vert \sum_{l=1}^{L}\pmb{V}^{(k)}(\beta_{(\Tilde{\mathcal{A}}_{2}^{k})^{c}})^{T}\hat{\pmb{W}}_{\tau_l}\pmb{\epsilon} \vert \vert_{\infty} \leq \sum_{l=1}^{L}\vert \vert \pmb{V}^{(k)}(\beta_{(\Tilde{\mathcal{A}}_{2}^{k})^{c}})^{T}\hat{\pmb{W}}_{\tau_l}\pmb{\epsilon} \vert \vert_{\infty} = O(n^{a/2+1/2}\sqrt{logn}).
\end{align*}
By condition 6, we could get
\begin{align*}
    (n\lambda_2)^{-1}\vert \vert \sum_{l=1}^{L}\pmb{V}^{(k)}(\beta_{(\Tilde{\mathcal{A}}_{2}^{k})^{c}})^{T}\hat{\pmb{W}}_{\tau_l}\pmb{\epsilon} \vert \vert_{\infty} = o(1).
\end{align*}
Next, we check the KKT condition of non-zero estimator.

For $j \in \mathcal{A}_{1}$, with probability tending to 1, 
\begin{align*}
    |\hat{\beta}_{j}| \geq |\beta_{j}^{0}| - |\hat{\beta}_{j} - \beta_{j}^{0}| \geq |\beta_{j}^{0}|- ||\pmb{\hat{\pmb{\theta}}}_{\mathcal{A}} -\pmb{\theta}^{0}_{\mathcal{A}} ||_{\infty} \geq |\beta_{j}^{0}| - ||\pmb{\hat{\pmb{\theta}}}_{\mathcal{A}} -\pmb{\theta}^{0}_{\mathcal{A}} ||_{2} > b_0 > a\lambda_1, 
\end{align*}
then $\min_{j \in \mathcal{A}_{1}} |\hat{\beta}_{j}| > a\lambda_1 $.

Similarly, we can also show that for each $k \in \lbrace 1,..,q \rbrace$, $\min_{j \in \mathcal{A}_{2}^{k}} |\gamma_{kj}| > a\lambda_2$ when $n$ is sufficiently large.

Therefore, we have shown that the corresponding KKT conditions are satisfied.
This completes the proof.

\section{Simulation}
In this section, we give an iterative coordinate descent algorithm for composite expectile regression. To demonstrate the performance of the proposed approach, we simulate two settings: homoscedastic setting and heteroscedastic setting. 
\subsection{Algorithm}
We use an iterative coordinate
descent (CD) algorithm, which optimizes the objective function (8) with respect to one of the four types of parameters $\pmb{b}_{l},\pmb{\alpha}, \pmb{\beta}, \pmb{\gamma}$. 
\begin{enumerate}
    \item Initialization: Let $t=0$, $\pmb{b}_{l}=0$, $\pmb{\beta}^{(t)} = 0$, $\pmb{\gamma}^{(t)} = 0$, $\pmb{\alpha}^{(t)}  = \pmb{(Z^{T}Z)^{-1}Z^{T}Y}$ for $l = 1,..,L$, and $\pmb{res}_{l}^{(t)} = \pmb{Y}-\pmb{b}_{l}^{(t)} - \pmb{Z}\pmb{\alpha}^{(t)}-\pmb{X}\pmb{\beta}^{(t)} - \sum_{k=1}^{q}\pmb{M}^{(k)}(\pmb{\beta}^{(t)}\odot \pmb{\gamma}_{k}^{(t)})$, where $\pmb{b}_{l}^{(t)}$, $\pmb{\alpha}^{(t)}, \pmb{\beta}^{(t)}, \pmb{\gamma}^{(t)}, \pmb{res}^{(t)}$ are the estimates of $\pmb{b}_{l}, \pmb{\alpha}, \pmb{\beta}, \pmb{\gamma}$ and residual vector at iteration $t$ respectively.
    \item Update $t = t + 1$. With fixed $\pmb{b}_l$, $\pmb{\gamma}$ and $\pmb{\alpha}$ at $\pmb{b}_{l}^{(t-1)}$, $\pmb{\gamma}^{(t-1)}$ and $\pmb{\alpha}^{(t-1)},l=1,...,L,$, we optimize $\overline{\mathcal{Q}}_{n}(\pmb{\theta})$ with respect to $\pmb{\beta}$. Let $\Tilde{\pmb{Y}}^{(t)}_{l} = \pmb{Y} - \pmb{b}_{l}^{(t-1)} - \pmb{Z}\pmb{\alpha}^{(t-1)}$ and $\Tilde{\pmb{X}}^{(t)} = \pmb{X} + \sum_{k=1}^{q}\pmb{M}^{(k)}\bigodot(\pmb{1}_{n\times 1}(\gamma_{k}^{(t-1)})^{T})$ with $\pmb{1}_{n\times 1} = (1,...,1)_{n\times 1}$. Then
    \begin{equation}
        \pmb{\beta}^{(t)} = \argmin \frac{1}{2n}\sum_{l=1}^{L}  ||\pmb{W}^{1/2}_{\tau_l}\left(\Tilde{\pmb{Y}}_{l}^{(t)} - \Tilde{\pmb{X}}^{(t)}\pmb{\beta} \right)||^2 + \sum_{j=1}^{p}\rho(|\beta_j|;\lambda_{1}, r)
    \end{equation}
    For $j = 1,..,p$, we run the following steps sequentially. $\pmb{W}_{\tau_l}$ is defined according to $\Tilde{\pmb{Y}}_{l}^{(t)} - \Tilde{\pmb{X}}^{(t)}\pmb{\beta}$ in each optimization steps with two possible elements $\tau_l, 1- \tau_l$. $\pmb{W}_{\tau_l}$ is defined differently under varied squared loss function. 
    
\begin{enumerate}
    \item Compute $ \pmb{res}_{-j,l}^{(t)} =\Tilde{\pmb{Y}}_{l}^{(t)}-\sum_{l=1}^{j-1}\Tilde{\pmb{X}}_{l}^{(t)}\beta_{l}^{(t)} -\sum_{l=j+1}^{p}\Tilde{\pmb{X}}_{l}^{(t)}\beta_{l}^{(t-1)}=\pmb{res}_{l}^{(t-1)} + \Tilde{\pmb{X}}_{j}^{(t)}\beta^{(t-1)}_{j}$,\\ $\phi_{j}^{(t)} =\frac{1}{n}\sum_{l=1}^{L}\left(\Tilde{\pmb{X}}_{j}^{(t)} \right)^{T}\pmb{W}_{\tau_{l}}\pmb{res}_{-j,l}^{(t)}$, $\psi_{j}^{(t)} = \frac{1}{n}\sum_{l=1}^{L}\left(\Tilde{\pmb{X}}_{j}^{(t)} \right)^{T}\pmb{W}_{\tau_{l}}\Tilde{\pmb{X}}_{j}^{(t)}$, $\pmb{W}_{\tau_l}$ is $n \times n $ diagonal matrix with two possible elements $\tau_l, 1- \tau_l$ and $\pmb{W}_{\tau_{l}}$ is updated at each step.
        
   \item update the estimate of $\beta_j$ as 
    \begin{equation}
\beta_{j}^{(t)} = \left\{
\begin{aligned}
&\frac{ST(\phi_{j}^{(t)}, \lambda_{1})}{\psi_{j}^{(t)} - \frac{1}{r}}, &|\phi_{j}^{(t)}| \leq \lambda_{1}r\psi_{j}^{(t)} \\
&\frac{\phi_{j}^{(t)}}{\psi_{j}^{(t)}}, & |\phi_{j}^{(t)}| > \lambda_{1}r\psi_{j}^{(t)}.
\end{aligned} \right.
\end{equation}
where $ST(v, \lambda_{1}) = sgn(v)(|v| - \lambda_{1})_{+}$ is the soft-thresholoding operator.

\item Update $\pmb{res}_{l}^{(t-1)} = \pmb{res}_{l}^{(t-1)} + \Tilde{\pmb{X}}_{j}^{(t)}\beta_{j}^{(t-1)} - \Tilde{\pmb{X}}_{j}^{(t)}\beta_{j}^{(t)}$
\end{enumerate}
    
\item With fixed $\pmb{b}_{l}$, $\pmb{\beta}$ and $\pmb{\alpha}$ at $\pmb{b}_{l}^{(t-1)}$, $\pmb{\beta}^{(t)}$ and $\pmb{\alpha}^{(t-1)}$, optimize equation (10) with respect to $\pmb{\gamma}$. Let $\hat{\pmb{Y}}_{l}^{(t)} = \pmb{Y} - \pmb{b}_{l}^{(t-1)} - \pmb{Z}\pmb{\alpha}^{(t-1)} - \pmb{X}\pmb{\beta}^{(t)}$ and $\left( \Tilde{\pmb{M}}^{(i)} \right)^{(t)} = \pmb{M}^{(i)}\bigodot \left( \pmb{1}_{n\times1}(\pmb{\beta}^{(t)})^{T}\right)$. Then
$$
\left( \gamma_{1}^{(t)},..., \gamma_{q}^{(t)} \right) = 
\argmin \frac{1}{2n}\sum_{l=1}^{L}   
||\pmb{W}^{1/2}_{\tau_l}\left(\hat{\pmb{Y}}_{
l}^{(t)}-\sum_{k=1}^{q}\left( 
\Tilde{\pmb{M}}^{(k)} 
\right)^{(t)}\gamma_{k}\right)||^{2}_{2} + 
\sum_{j=1}^{p}\sum_{k=1}^{q}\rho(|\gamma_{kj}|;\lambda_{2
},r)
$$
For $k= 1,...,q$ and $j \in \lbrace j: \beta_{j}^{(t)} \neq 0, j = 1,...,p \rbrace$, conduct estimation similar to step 2.
\item With fixed $\pmb{b}_{l}$, $\pmb{\beta}$ and $\pmb{\gamma}$ at $\pmb{b}_{l}^{(t-1)}$,  $\pmb{\beta}^{(t)}$ and $\pmb{\gamma}^{(t)}$, we optimize equation (10) with respect to $\pmb{\alpha}$. Let $\hat{\pmb{Y}}_{l}^{(t)} = \pmb{Y}-\pmb{b}_{l}^{(t-1)}-\pmb{X}\pmb{\beta}^{(t)} - \sum_{k=1}^{q}\pmb{M}^{(k)}(\pmb{\beta}^{(t)}\odot \pmb{\gamma}_{k}^{(t)})$. Then
$$
\pmb{\alpha}^{(t)} = \argmin
\frac{1}{2n}\sum_{l=1}^{L}||\pmb{W}^{1/2}_{\tau_l}\left(\
hat{\pmb{Y}}_{l}^{(t)}-\pmb{Z\alpha}\right)||^{2}_{2}
$$
We can get $\pmb{\alpha}^{(t)} = (\sum_{l=1}^{L}\pmb{Z}^{T}\pmb{W}_{\tau_l}\pmb{Z})^{-1}(\sum_{l=1}^{L}\pmb{Z}^{T}\pmb{W}_{\tau_l}\hat{\pmb{Y}}_{l}^{(t)})$, where $\hat{\pmb{Y}}_{l}^{(t)} = \pmb{res}_{l}^{(t-1)} + \pmb{Z}\pmb{\alpha}^{(t-1)}$. Then we update $\pmb{res}_{l}^{(t)} = \pmb{res}_{l}^{(t-1)} + \pmb{Z}\pmb{\alpha}^{(t-1)} - \pmb{Z}\pmb{\alpha}^{(t)}$ for each $l$ from 1 to $L$.
\item With fixed $\pmb{\alpha}$, $\pmb{\beta}$ and $\pmb{\gamma}$ at $\pmb{\alpha}^{(t)}$,  $\pmb{\beta}^{(t)}$ and $\pmb{\gamma}^{(t)}$,  we optimize equation (10) with respect to $\pmb{b}_{l}$. Let $\hat{\pmb{Y}}^{(t)} = \pmb{Y}-\pmb{Z}\pmb{\alpha}^{(t)}-\pmb{X}\pmb{\beta}^{(t)} - \sum_{k=1}^{q}\pmb{M}^{(k)}(\pmb{\beta}^{(t)}\odot \pmb{\gamma}_{k}^{(t)})$. Then
$$
(\pmb{b}_1,...,\pmb{b}_{L}) = \argmin \frac{1}{2n}\sum_{l=1}^{L}||\pmb{W}^{1/2}_{\tau_l
}\left(\hat{\pmb{Y}}^{(t)}-\pmb{b}_{l}\right)||^{
2}_{2}
$$
This is a convex optimization problem with respect to $\pmb{b}_l$. A build-in function 'optimise' in R is used to get the solution. Then we update $\pmb{res}_{l}^{(t)} = \pmb{res}_{l}^{(t-1)} + \pmb{b}_{l}^{(t-1)} - \pmb{b}_{l}^{(t)}$.
\item Repeat step 2 to step 5 until convergence: $$\frac{|\overline{\mathcal{Q}}_{n}(\pmb{\theta}^{(t)}) - \overline{\mathcal{Q}}_{n}(\pmb{\theta}^{(t-1)})|}{|\overline{\mathcal{Q}}_{n}(\pmb{\theta}^{(t-1)})|} < 10^{-4}$$
\end{enumerate}

\subsubsection{Details of step 2 for proposed algorithm}
For $j = 1, . . . , p$, the CD algorithm optimizes the objective function
with respect to $\beta_j$ while fixing the other parameters $\beta_{l}(l \neq j)$ at 
their current estimates 
$\beta_{l}^{(t)}$ for $l<j$ or $\beta_{l}^{(t-1)}$ for $l>j$.
Consider the following simplified objective function in step 2:
\begin{equation}
    \overline{\mathcal{Q}}_{s}(\beta_{j}) = \frac{1}{2n}\sum_{l=1}^{L} ||\pmb{W}_{\tau_{l}}^{1/2}(\pmb{res}_{-j,l}^{(t)}-\Tilde{\pmb{X}}_{j}^{(t)}\beta_{j})||_{2}^{2} + \rho(|\beta_{j}|;\lambda_1, r) 
\end{equation}
    where $\pmb{res}_{-j,l}^{(t)} = \Tilde{\pmb{Y}}_{l}^{(t)} - \sum_{l=1}^{j-1}\Tilde{\pmb{X}}_{l}^{(t)}\beta_{l}^{(t)} -\sum_{l=j+1}^{p}\Tilde{\pmb{X}}_{l}^{t}\beta_{l}^{(t-1)} = \pmb{res}_{l}^{(t-1)} + \Tilde{\pmb{X}}_{j}^{(t)}\beta_{j}^{(t-1)}$
with $\pmb{res}_{-j,l}^{(t)} = \Tilde{\pmb{Y}}_{l}^{(t)} - \sum_{l=1}^{j-1}\Tilde{\pmb{X}}_{l}^{(t)}\beta_{l}^{(t)} - \sum_{l=j}^{p}\Tilde{\pmb{X}}_{l}^{t}\beta_{l}^{(t-1)}$, $\pmb{W}_{\tau_l}$ is $n \times n $ diagonal matrix with two possible elements $\tau_l, 1- \tau_l$

We take derivative with respective to $\beta_{j}$ 
\begin{align*}
 \frac{\partial\overline{\mathcal{Q}}_{s}(\mathbf{\beta})}{\partial \beta_j}
 &=-\frac{1}{n}\sum_{l=1}^{L}\left(\Tilde{\pmb{X}}_{j}^{(t)} \right)^{T}\pmb{W}_{\tau_{l}}\pmb{res}_{-j,l}^{(t)} + \frac{1}{n}\sum_{l=1}^{L}\left(\Tilde{\pmb{X}}_{j}^{(t)} \right)^{T}\pmb{W}_{\tau_{l}}\Tilde{\pmb{X}}_{j}^{(t)}\beta_{j}\\
 &+\lambda_{1}sgn(\beta_{j}) \left\{
\begin{aligned}
&1 - \frac{|\beta_{j}|}{\lambda_{1}r}, & |\beta_{j}| \leq \lambda_{1}r \\
& 0, & |\beta_{j}| > \lambda_{1}r
\end{aligned}
\right. \\
& = -\phi_{j}^{(t)} + \psi_{j}^{(t)}\beta_j + \lambda_{1}sgn(\beta_{j}) \left\{
\begin{aligned}
&1 - \frac{|\beta_{j}|}{\lambda_{1}r}, & |\beta_{j}| \leq \lambda_{1}r \\
& 0, & |\beta_{j}| > \lambda_{1}r
\end{aligned}
\right.
\end{align*}
where $\phi_{j}^{(t)} 
=\frac{1}{n}\sum_{l=1}^{L}\left(\Tilde{\pmb{X}}_{j}^{
(t)} \right)^{T}\pmb{W}_{\tau_{l}}\pmb{res}_{-j,l}^{(
t)}$, $\psi_{j}^{(t)} = 
\frac{1}{n}\sum_{l=1}^{L}\left(\Tilde{\pmb{X}}_{j}^{(
t)} \right)^{T}\pmb{W}_{\tau_{l}}\Tilde{\pmb{X}}_{j}^
{(t)}$.

By setting $ \frac{\partial\overline{\mathcal{Q}}_{s}
(\mathbf{\beta})}{\partial \beta_j} = 0$, we have
\begin{equation}
\beta_{j}^{(t)} = \left\{
\begin{aligned}
&\frac{ST(\phi_{j}^{(t)}, 
\lambda_{1})}{\psi_{j}^{(t)} - \frac{1}{r}}, 
&|\phi_{j}^{(t)}| \leq \lambda_{1}r\psi_{j}^{(t)} \\
&\frac{\phi_{j}^{(t)}}{\psi_{j}^{(t)}}, & 
|\phi_{j}^{(t)}| > \lambda_{1}r\psi_{j}^{(t)}.
\end{aligned} \right.
\end{equation}
where $ST(v, \lambda_{1}) = sgn(v)(|v| - 
\lambda_{1})_{+}$ is the soft-thresholoding operator.
\subsection{Simulation setting}
We choose n = 500, q = 5 and p = 5,00. Thus, there 
are a total of 5,05 main effects and 25,00 
interactions. The true genetic estimators and G-E interaction estimators are fixed.
\begin{itemize}
    \item For G effect, we simulate gene expression data coded  from a multivariate Normal distribution.
    \item For E factors, we first generate five continuous variables from a multivariate Normal distribution with marginal mean 0, marginal variance 1, and correlation structure AR(0.3), and then dichotomize two of them at 0 to create two binary variables. There are thus three continuous and two binary E factors.
    \item For E factors, their coefficients $\alpha$ are generated from Uniform (0.8, 1.2) with 5 non-zero value.
    \item For G factors, their coefficients $\beta$ are filled in sparse form, with 20 non-zero value.
    \item For G-E interaction effects, their coefficients $\gamma_{kj}$ are also filled in sparse uniform, with 40 non-zero value.  
    \item We simulate y as a continuous response based on model (1)
\end{itemize}
To find optimal tuning parameters($\lambda_1$,$\lambda_2$, $r$), we use grid search method through a specified subset. For example, $\lambda_1=(0.1,0.5,1,1.5,2),\lambda_2=(0.1,0.5,1,1.5,2)$ and $r = 3$. 1000 iterations are used to train both ER and CER. To measure the model performance, we define absolute estimation error(AE) and Square estimation error(SE)
\begin{itemize}
    \item Absolute estimation error:\\ AE=$(\sum_{k=1}^{q}|\hat{\alpha}_{k} -\hat{\alpha}^{0}_{k}| + \sum_{j=1}^{p}|\hat{\beta}_{j} - \beta^{0}_{j}| + \sum_{k=1}^{q}\sum_{j=1}^{p}|
    \hat{\gamma}_{kj}-\gamma_{kj}^{0}|)$ 
    \item Square estimation error:\\ SE = $\sqrt{(\sum_{k=1}^{q}(\hat{\alpha}_{k} -\hat{\alpha}^{0}_{k})^{2} + \sum_{j=1}^{p}(\hat{\beta}_{j} - \beta^{0}_{j})^2 + \sum_{k=1}^{q}\sum_{j=1}^{p}(\hat{\gamma}_{kj}-\gamma_{kj}^{0})^{2})}$
    \item Mean Absolute Deviation:\\ MAD = $\frac{1}{nL}\sum_{l=1}^{L}\mid \mid(\pmb{Y}-\hat{\pmb{b}}_{l} - \pmb{Z\hat{\alpha}}-\pmb{X}_{\mathcal{A}_1}\pmb{\hat{\beta}}_{\mathcal{A}_{1}} - \sum_{k=1}^{q}\pmb{M}^{(k)}_{\mathcal{A}_{2}^{k}}(\pmb{\hat{\beta}}_{\mathcal{A}_{2}^{k}}\odot \pmb{\hat{\gamma}}_{k,\mathcal{A}_{2}^{k}}))\mid \mid_{1}$
\end{itemize}
 as criterion to measure the model's estimation performance. True positive(TP) and False positive(FP) are also proposed to measure the model's selection performance. We use 
 $$BIC = \frac{C}{nL}\sum_{l=1}^{L}\mid \mid\pmb{W}^{1/2}_{\tau_{l}}(\pmb{Y}-\hat{\pmb{b}}_{l} - \pmb{Z\hat{\alpha}}-\pmb{X}_{\mathcal{A}_1}\pmb{\hat{\beta}}_{\mathcal{A}_{1}} - \sum_{k=1}^{q}\pmb{M}^{(k)}_{\mathcal{A}_{2}^{k}}(\pmb{\hat{\beta}}_{\mathcal{A}_{2}^{k}}\odot \pmb{\hat{\gamma}}_{k,\mathcal{A}_{2}^{k}}))\mid \mid^{2}_{2} + s\frac{C_{n}log(v)}{n}
 $$
 to select models,where C is a constant, $v= p*q+p+q,C_n= log(log(n)), s = DF$. This data-driven BIC criterion is motivated by Gu and Zou[10]. In composite expectile regression, we choose L as 9 or 19 to combine the strength of multiple expectile regressions. DF is the total number of non-zero estimator of G effect and G-E interaction effect.

Simulation studies were conducted to compare the performance of CER, ER under different 
settings: homoscedastic setting and heteroscedastic setting. Totally 200 replicates
were simulated for each simulation setting. In each replicate, we measure the model performance in term of selection and estimation in high dimensional statistics, 
\subsection{Setting I: homoscedastic setting}
In this section, we consider two types of error distribution:
\begin{enumerate}
    \item Normal distribution: $\epsilon\sim N(0,1)$
    \item Student's t distribution with degree of freedom 4: $\epsilon \sim \frac{1}{\sqrt{2}} t(4)$ 
\end{enumerate}
Since $\tau-$ expectile quantifies different "location" of a distribution, we use 5 different $\tau = 0.1, 0.25, 0.5, 0.75, 0.9$ to get a comprehensive view of relationship between covariate and response. If $\tau = 0.5$, expectile regression is degenerate to mean regression. Therefore, expectile regression can be regarded as a generalization of the mean and an alternative measure of "location" of a distribution (Gu and Zou 2020). 


\begin{table}[h]
\caption{Error distribution is normal model under 200 replicates}
\begin{tabular}{r|r|r|r|r|r|r|r|r|r|r|r|r}
\hline
Method & AE & SE   & TP  & FP &MAD\\
\hline
ER with $\tau=0.10$ & 22.38(6.52) & 3.44(0.89)  & 50.53(2.54) & 8.76(8.27) & 1.41(0.21)\\
\hline
ER with $\tau=0.25$ & 15.57(2.88) & 2.53(0.39)  & 52.98(1.75) & 7.12(6.71)& 1.17(0.11)\\
\hline
ER with $\tau=0.50$ & 14.35(2.88) & 2.38(0.36)  & 53.30(1.57) & 6.22(7.82) &1.14(0.11)\\
\hline
ER with $\tau=0.75$ & 15.65(2.78) & 2.57(0.39)  & 52.87(1.48) & 6.60(5.47) & 1.18(0.11)\\
\hline
ER with $\tau=0.90$ & 22.24(6.26) & 3.41(0.80)  & 50.76(2.28) & 9.79(8.92) & 1.39(0.19)\\
\hline
CER &10.69(2.02) & 1.69(0.35)  & 57.59(1.16) & 8.45(3.97)& 1.09(0.12)\\
\hline
non-hierarchical CER & 17.64(2.7) & 2.95(0.47)  & 55.20(2.08) & 8.49(3.51)  & 1.31(0.07)\\
\hline
\end{tabular}
\end{table}
Based on results from table 1, CER has smaller AE and SE value than ER and non-hierarchical CER under different $\tau$ value. Total 60 fixed G effect estimators and G-E interaction estimators are simulated. CER tends to identify more main G effects and G-E interaction effects while achieve comparable more TP. It also have comparable FP. The proposed approach has better prediction performance in term of MAD.

\begin{table}[h]
\caption{Error distribution is t/$\sqrt{2}$ distribution under 200 replicates}
\begin{tabular}{r|r|r|r|r|r|r|r|r|r|r|r|r}
\hline
Method & AE & SE  & TP  & FP &MAD\\
\hline
ER with $\tau=0.10$ & 23.43(6.40) & 3.58(0.85)  & 50.24(2.68) & 9.61(8.27)&1.44(0.24)\\
\hline
ER with $\tau=0.25$ & 16.49(3.67) & 2.68(0.46)  & 52.49(1.55) & 7.28(8.11) & 1.16(0.14)\\
\hline
ER with $\tau=0.50$ & 14.89(3.34) & 2.46(0.41)  & 52.99(1.45) & 8.70(10.97) & 1.07(0.15)\\
\hline
ER with $\tau=0.75$ & 16.65(3.56) & 2.69(0.45)  & 52.31(1.86) & 8.24(8.85) & 1.17(0.13)\\
\hline
ER with $\tau=0.90$ & 23.82(7.35) & 3.62(0.96)  & 50.11(2.76) & 11.38(11.42) & 1.42(0.21)\\
\hline
CER & 9.63(1.74) & 1.53(0.30)  & 57.73(1.07) & 6.22(2.48) & 1.01(0.12)\\
\hline
non-hierarchical CER & 16.93(2.76) & 2.94(0.52)  & 55.65(2.04) & 10.43 (4.66)  & 1.23(0.10)\\
\hline
\end{tabular}
\end{table}

In table 2, we have similar result in table 1. CER outperforms ER and non-hierarchical CER in term of AE and SE. The proposed method also has better identification performance to choose more fixed G-effect estimators and G-E interaction estimators. In the meantime, the number of false positives is smaller.  MAD of CER are comparable to ER.

\subsection{Setting II: Heteroscedastic setting}

We adopt a model from Wang, Wu and Li (2012). In the model,
the covariates are generated in two steps. First, we generate copies of $\pmb{Z}_{n\times q}=(\pmb{Z}^{c}_{1},..., \pmb{Z}^{c}_q)$
and $\pmb{X}_{n\times p} = (\pmb{X}^{c}_{1},...,\pmb{X}^{c}_{p})$. In the
second step, for each copy of $(\pmb{Z}^{c}_1,..., \pmb{Z}^{c}_{q})$ and $(\pmb{X}^{c}_{1},...,\pmb{X}^{c}_{p})$, we set $\pmb{Z}_1 = \pmb{\Phi}(\pmb{Z_1}^{c})$, $\pmb{Z}_j = \pmb{Z}^{c}_j$, $\pmb{X}_1 = \pmb{\Phi}(\pmb{X_1}^{c})$ and $\pmb{X}_j = \pmb{X}^{c}_j$ for
$j = 2,3,...,q$, where $\pmb{\Phi(\cdot)}$ is the standard normal CDF. 
To include heteroscedastic error, we generate the data in this form:
\begin{align}
  Y_i = \pmb{Z_{i}}\pmb{\alpha}+\pmb{X_{i}}\pmb{\beta}+\sum_{k=1}^{q}\pmb{M_{i}}^{(k)}(\pmb{\beta}\odot \pmb{\gamma}_{k})+|\sigma(\pmb{Z_{i}}, \pmb{X_{i}})|\epsilon_{i}
\end{align}
$|\sigma(\pmb{Z_{i}}, \pmb{X_{i}})| = \pmb{Z}_1 + \pmb{X}_{1}$,
 $\epsilon_i \sim N(0,1)$.

\begin{table}[h]
\caption{Error distribution is heteroscedastic under 200 replicates }
\begin{tabular}{r|r|r|r|r|r|r|r}
\hline
Method & AE & SE & TP & FP &MAD\\
\hline

ER with $\tau=0.10$ & 19.96(6.84) & 3.12(0.90)  & 51.04(2.71) & 8.65(9.77) & 1.17(0.29)\\
\hline
ER with $\tau=0.25$ & 14.32(3.29) & 2.42(0.46)  & 52.74(1.49) & 5.66(7.74)& 0.92(0.13)\\
\hline
ER with $\tau=0.50$ & 12.92(2.63) & 2.23(0.31)  & 53.48(1.30) & 6.59(8.79) & 0.85(0.10)\\
\hline
ER with $\tau=0.75$ & 14.13(2.60) & 2.41(0.35) & 52.81(1.35) & 5.22(6.17) & 0.92(0.12)\\
\hline
ER with $\tau=0.90$ & 20.36(6.65) & 3.19(0.89)  & 50.85(2.77) & 9.12(9.73)  & 1.17(0.27)\\
\hline
CER & 7.77(1.36) & 1.48(0.3)  & 58.21(0.85) & 3.21(1.94)& 0.44(0.11)\\
\hline
non-hierarchical CER &12.38(3.71) & 2.32(0.57)  & 57.35(2.25) & 4.78(3.54) & 0.65(0.2)\\
\hline
\end{tabular}
\end{table}
From table 3, the proposed approach has better estimation and selection performance than ER with different $\tau$ levels and CER without hierarchy.

\section{Real data analysis}
Lung adenocarcinoma(LUAD) occurs due to abnormal and uncontrolled cell growth in the lungs, which is a subtype of non-small cell lung cancer that is often diagnosed in an outer area of the lung(\url{https://rarediseases.info.nih.gov/diseases/5742/lung-adenocarcinoma}). LUAD evolves from the mucosal glands and represents about 40\% of all lung cancers. This rare disease is the most common disease to be diagnosed in people who have never smoked. LUAD usually occurs in the lung periphery, and in many cases, may be found in scars or areas of chronic inflammation(Myers and
Wallen, 2021).

In this section, we applied composite expectile regression as well as the 
alternative to lung adenocarcinoma. We select age, gender, patient smoking history, and 
pathologic tumor stage as environmental variables, all of which have been 
suggested to be potentially associated with LUAD(Du, Liu and Wu(2021)). There are 522 subjects in this 
study. The FEV1 is response which measures how much air you can exhale in one second and can be used in the diagnosis of obstructive and restrictive lung disease. We 
match the mRNA gene expression measurements with the clinical/environmental 
variables and response. For genetic effects, after some preprocessing procedures 
including matching subjects and imputing missing data, 232 subjects and 20097 mRNA measurements are chosen. Here, we select 791 mRNA with marginal screening.
We examine prediction performance using a 
resampling-based approach. Specifically, 
subjects are randomly split into a 
training and a testing set with a ratio of
7:3. Then we estimate parameters using the
training set and make prediction for the 
testing set subjects. With 50 resamplings, we compute the mean of all MADs.
 CER with MAD = 0.823. ER with MAD=0.857($\tau = 0.5$), MAD = 0.858($\tau=0.25$), MAD = 0.868($\tau=0.75$), MAD = 0.889($\tau=0.1$), MAD = 0.885($\tau=0.9$)

 The proposed approach identified genes with implications of LUAD.  281 of  main effects and interactions are identified. The details of coefficients can be found in appendix, table 4.
 The current study demonstrated that AATK, a radiosensitization-associated gene, is a target of miR‑558 in lung cancer cells, using in silico analysis and a luciferase reporter system(Zhu ect,2016). High gene ABCG4 expression is associated with poor prognosis in non-small-cell lung cancer patients treated with cisplatin-based chemotherapy(Yang ect, 2015). Gene ADAMTS20 mutations and high amplification of NKX2-1 may be related to brain metastases of lung cancer(Li ect, 2020). Gene AGFG1 is prognostic, high expression is unfavorable in lung cancer, which is  associated with severity of airway responsiveness(Himes, 2013). Gene ALG13 were significantly associated with lymph node status of patients with  non-small-cell lung cancer(Deng ect, 2019). Gene ARL2 induces
rapid release of deltarasin from phosphodiesterase 6 delta subunit, resulting in the impairment of KRAS-dependent lung cancer cell growth(Leung ect, 2018). ATG16L1 is associated with decreased risk of brain metastasis in patients with non-small cell lung cancer(Li ect, 2017). Targeted gene BMI1 inhibition impairs tumor growth in lung adenocarcinomas with low CEBP$\alpha$ expression(Yong ect, 2016).


\section{Summary and discussion}
In this article, we have studied the sparse penalized CER with G-E interaction. In particular, we establish the selection and estimation consistency of the CER esimator. By implementing coordinate descent algorithm, we have shown that CER has superior or comparable model performance, compared to alternatives. Under heteroscedastic setting, simulations show that CER has better selection and estimation performance. A real data is analyzed to demonstrate the performance of proposed approach.

For CER, it is computational expensive compared to expectile regression since we integrate multiple expectile regression in the loss function. It is worthwhile to explore if an efficient algorithm could be implemented to speed up the computation. We investigate the selection and estimation consistency of CER. It is natural to explore the normality of CER in the future under high dimensional statistics. Some theoretical works has been done in quantile regression scenario. For example, if $L$ goes to infinity,  the estimators by combining quantile regressions are asymptotically efficient. We fixed L in our loss function, it is also worth to investigate the theoretical properties if $L$ goes to infinity. 
%
\section{Appendix}

\begin{table}[]
\caption{Analysis of the LUAD data using CER: identified main effects and interactions}
\begin{tabular}{|l|l|l|l|l|l|}

\hline
Gene     &               & Sex          & Patient Smoking History& Diagnosis Age & Cancer.Tumor.Stage \\ \hline
     &               & 0.0616   & 0.1235                       & -0.0216   & 0.1120                                          \\ \hline
VLOC645851   & -0.4068  &              & 0.0434                      & 0.0060   & 0.0587                                         \\ \hline
TMPRSS11E2   & -0.5038  & -0.0106 & 0.0418                      & 0.0071   & -0.0141                                        \\ \hline
AATK   & -0.0135  &              &                                  & -0.0022  &                                                     \\ \hline
ABCG4   & 0.0156  &              &                                  & 0.0038   &                                                     \\ \hline
ASIC4   & -0.0206  &              &                                  & -0.0036  &                                                     \\ \hline
ADAM3A  & -0.0074  &              &                                  & -0.0025  &                                                     \\ \hline
ADAMTS1  & -0.0079  &              &                                  & -0.0040  &                                                     \\ \hline
ADAMTS20  & -0.0012  &              &                                  &               &                                                     \\ \hline
AGFG1  & -0.7800   & 0.1285   &                                  & 0.0117    & 0.1708                                         \\ \hline
AKR1D1  & -0.0137 &              &                                  & -0.0039  &                                                     \\ \hline
ALG10  & 0.0117   &              &                                  & 0.00190   &                                                     \\ \hline
ALG13  & -0.3321  &              & -0.0131                     & 0.0097   &                                                     \\ \hline
ANKRD55  & -0.5915  &              & 0.0706                    & -0.0036  & 0.2206                                         \\ \hline
AQP8  & -0.0088  &              &                                  & -0.0012  &                                                     \\ \hline
ARL2  & -0.0082  &              &                                  & -0.0023  &                                                     \\ \hline
ATG16L1  & -0.0176  &              &                                  & -0.0045  &                                                     \\ \hline
BMI1  & -0.0017   &              &                                  &               &                                                     \\ \hline
BMX  & 0.6888    &              & -0.0279                    & -0.0136   & -0.0446                                         \\ \hline
DEPP1  & 0.0112  &              &                                  & 0.0037   &                                                     \\ \hline
DDIAS  & 0.0182  &              &                                  & 0.0072   &                                                     \\ \hline
GPATCH2L  & -0.0294  &              &                                  & -0.0029  &                                                     \\ \hline
C14orf178  & 0.0050  &              &                                  & 0.0004  &                                                     \\ \hline

\end{tabular}
\end{table}

\begin{table}[]
\caption{Continued analysis of the LUAD data using CER: identified main effects and interactions}
\begin{tabular}{|l|l|l|l|l|l|}
\hline
Gene     &               & Sex          & Patient Smoking History& Diagnosis Age & Cancer.Tumor.Stage \\ \hline
SLC25A47  & 0.0116    &              &                                  & 0.0022    &                                                     \\ \hline
C16orf92  & 0.0162   &              &                                  & 0.0012  &                                                     \\ \hline
SPATA46  & -0.0025 &              &                                  &               &                                                     \\ \hline
METTL18  & -0.0117&              &                                  & -0.0008 &                                                     \\ \hline
C3orf49  & -0.0117 &              &                                  & -0.0015  &                                                     \\ \hline
NADK2  & -0.0141  &              &                                  & -0.0035  &                                                     \\ \hline
FAM225B  & -0.0082  &              &                                  & -0.0039  &                                                     \\ \hline
CARD18  & -0.0063 &              &                                  & -0.0001 &                                                     \\ \hline
CATSPER2P1 & -0.0107 &              &                                  & -0.0005 &                                                     \\ \hline
CCDC163P & -0.0166  &              &                                  & -0.0031  &                                                     \\ \hline
CCDC91 & -0.0124  &              &                                  & -0.0046  &                                                     \\ \hline
ACKR4 & -0.0141  &              &                                  & -0.0020  &                                                     \\ \hline
CD36 & -0.0074  &              &                                  & -0.0051  &                                                     \\ \hline
CDC42SE1 & -0.0208  &              &                                  & -0.0062  &                                                     \\ \hline
CEP72 & -0.0103  &              &                                  & -0.0035 &                                                     \\ \hline
CLCN4 & 0.0102  &              &                                  & 0.0010  &                                                     \\ \hline
CLDN10 & 0.0131  &              &                                  & 0.0010  &                                                     \\ \hline
CLN3 & -0.0094 &              &                                  & -0.0039  &                                                     \\ \hline
COLEC10 & -0.0001 &              &                                  &               &                                                     \\ \hline
CRISPLD1 & -0.1900  &              & 0.0309                      & -0.0014  &                                                     \\ \hline
CSF3R & 0.0191   &              &                                  & 0.0042   &                                                     \\ \hline
CYMP & -0.0107   &              &                                  & -0.0034   &                                                     \\ \hline

\end{tabular}
\end{table}

\begin{table}[]
\caption{Continued analysis of the LUAD data using CER: identified main effects and interactions}
\begin{tabular}{|l|l|l|l|l|l|}
\hline
Gene     &               & Sex          & Patient Smoking History& Diagnosis Age & Cancer.Tumor.Stage \\ \hline

CYP27C1 & -0.0163  &              &                                  & -0.0023  &                                                     \\ \hline
                                                      
DCAF16 & -0.0128  &              &                                  & -0.0030  &                                                     \\ \hline
DEPDC4 & 0.0113   &              &                                  & 0.0018   &                                                     \\ \hline
DGKG & 0.0348   &              &                                  & 0.0024   &                                                     \\ \hline
DHRS4 & 0.0065   &              &                                  &               &                                                     \\ \hline
DSPP & 0.0061  &              &                                  & 0.0010  &                                                     \\ \hline
EAF2 & -0.0024  &              &                                  &               &                                                     \\ \hline
ECM1 & 0.0416   &              &                                  & 0.0011  &                                                     \\ \hline
EIF1AD & -0.0241  &              &                                  & -0.0043  &                                                     \\ \hline
ETV5 & 0.0070  &              &                                  & 0.0007  &                                                     \\ \hline
FAH & 0.0042   &              &                                  &               &                                                     \\ \hline
FAM48B2 & 0.0324   &              &                                  & 0.0054   &                                                     \\ \hline
FIP1L1 & 0.0709   &              &                                  & 0.0053  &                                                     \\ \hline
FOS & 0.0186   &              &                                  & 0.0039   &                                                     \\ \hline
GALNT13 & -0.0106 &              &                                  & -0.0010 &                                                     \\ \hline
FLJ21230 & 0.0375   &              &                                  & 0.0016   &                                                     \\ \hline
GPD1 & -0.0014  &              &                                  &               &                                                     \\ \hline
GRHL3 & 0.0101   &              &                                  & 0.0046   &                                                     \\ \hline
GSTT2 & -0.0256  &              &                                  & -0.0013  &                                                     \\ \hline
HEPN1 & -0.0429  &              & -0.0047                    & -0.0053  &                                                     \\ \hline
HMGN4 & -0.0145  &              &                                  & -0.0055  &                                                     \\ \hline
HMHB1 & 0.0444   &              & 0.0168                   & 0.0023   &                                                     \\ \hline

\end{tabular}
\end{table}

\begin{table}[]
\caption{Continued analysis of the LUAD data using CER: identified main effects and interactions}
\begin{tabular}{|l|l|l|l|l|l|}
\hline
Gene     &               & Sex          & Patient Smoking History& Diagnosis Age & Cancer.Tumor.Stage \\ \hline

HNRNPR & -0.0601  &              &                                  & -0.0072  &                                                     \\ \hline
HPR & -0.0252  &              &                                  & -0.0017  &                                                     \\ \hline
HSD3B2 & 0.01560   &              &                                  & 0.0056   &                                                     \\ \hline
IL31RA & -0.0115  &              &                                  & -0.0013  &                                                     \\ \hline
IWS1 & 0.0022   &              &                                  &               &                                                     \\ \hline
KCNA5 & 0.0088   &              &                                  & 0.0047   &                                                     \\ \hline
KCNT1 & 0.0089   &              &                                  & 0.00113   &                                                     \\ \hline
KDM2A & 0.0185   &              &                                  & 0.0037   &                                                     \\ \hline
ATG13 & -0.0117  &              &                                  & -0.0044  &                                                     \\ \hline
ICE1 & 0.0481   &              &                                  & 0.0037   &                                                     \\ \hline
KIF5A & 0.0136  &              &                                  & 0.0040   &                                                     \\ \hline
KLC3 & -0.0146  &              &                                  & -0.0023  &                                                     \\ \hline
KRT72 & 0.0040   &              &                                  &               &                                                     \\ \hline
KRTAP4-11 & -0.026  &              &                                  & -0.0076  &                                                     \\ \hline
LGALS9C & -0.0352 &              &                                  & -0.0016  &                                                     \\ \hline
EPT & 0.0225   &              &                                  & 0.0083   &                                                     \\ \hline
LHX9 & -0.0141  &              &                                  & -0.0055  &                                                     \\ \hline
LIPL4 & 0.0192   &              &                                  & 0.0035   &                                                     \\ \hline
LOC100240726 & 0.0799    &              &                                  & -0.0063   &                                                     \\ \hline
PACRG-AS1 & 0.0024    &              &                                  &               &                                                     \\ \hline
LY6E & 0.0186   &              &                                  & 0.0036   &                                                     \\ \hline
LYPD2 & 0.0154  &              &                                  & 0.0007  &                                                     \\ \hline

\end{tabular}
\end{table}

\begin{table}[]
\caption{Continued analysis of the LUAD data using CER: identified main effects and interactions}
\begin{tabular}{|l|l|l|l|l|l|}
\hline
Gene     &               & Sex          & Patient Smoking History& Diagnosis Age & Cancer.Tumor.Stage \\ \hline

EEF1AKNMT & -0.0131  &              &                                  & -0.0019  &                                                     \\ \hline
STEAP1B & -0.0091 &              &                                  & -0.0007 &                                                     \\ \hline
MKRN2 & -0.0146  &              &                                  & -0.0027  &                                                     \\ \hline
                                                      
DLG6 & -0.0071  &              &                                  & -0.0042  &                                                     \\ \hline
NAP1L2 & -0.0162  &              &                                  & -0.0056  &                                                     \\ \hline
LINC00032 & 0.0102   &              &                                  & 0.0019   &                                                     \\ \hline
LINC00114 & 0.0007   &              &                                  &               &                                                     \\ \hline
ANKRD30BL & 0.0147   &              &                                  & 0.0045  &                                                     \\ \hline
NNMT & 0.0142  &              &                                  & 0.0060   &                                                     \\ \hline
NOL11 & 0.0215   &              &                                  & 0.0018   &                                                     \\ \hline
NPAS2 & 0.0117  &              &                                  & 0.0008  &                                                     \\ \hline
FLJ22583 & -0.0180  &              &                                  & -0.0018  &                                                     \\ \hline
OMG & -0.0121   &              &                                  & -0.0016   &                                                     \\ \hline
OR2A17P & -0.0081  &              &                                  & -0.0027  &                                                     \\ \hline
OR4C6 & 0.0177   &              &                                  & 0.0027   &                                                     \\ \hline
OR4D1 & 0.0058   &              &                                  & 0.0005   &                                                     \\ \hline
ORMDL1 & -0.0110  &              &                                  & -0.0017   &                                                     \\ \hline
OVCH2 & 0.0170  &              &                                  & 0.0008  &                                                     \\ \hline
OXCT2 & -0.0537  &              & -0.021602955                     & -0.0013  &                                                     \\ \hline
CNRS7 & 0.0275  &              &                                  & 0.0007  &                                                     \\ \hline
PCDHA2 & 0.0411   &              &                                  & 0.0010  &                                                     \\ \hline
PDIK1L & 0.0075   &              &                                  & 0.0029  &                                                     \\ \hline

\end{tabular}
\end{table}

\begin{table}[]
\caption{Continued analysis of the LUAD data using CER: identified main effects and interactions}
\begin{tabular}{|l|l|l|l|l|l|}
\hline
Gene     &               & Sex          & Patient Smoking History& Diagnosis Age & Cancer.Tumor.Stage \\ \hline

PF4 & 0.0023    &              &                                  &               &                                                     \\ \hline
PGK2 & -0.0077  &              &                                  & -0.0042  &                                                     \\ \hline
PHF7 & 0.0172   &              &                                  & 0.0022   &                                                     \\ \hline
PIGQ & 0.0171   &              &                                  & 0.0016   &                                                     \\ \hline
PREP & 0.0086   &              &                                  & 0.0031   &                                                     \\ \hline
PSAPL1 & 0.0149  &              &                                  & 0.0007  &                                                     \\ \hline
RAB20 & -0.0090  &              &                                  & -0.0065  &                                                     \\ \hline
RASL11A & 0.0112   &              &                                  & 0.0010  &                                                     \\ \hline
RFPL1 & -0.0138  &              &                                  & -0.0033  &                                                     \\ \hline
RGPD5 & -0.0084  &              &                                  & -0.0058  &                                                     \\ \hline
RIMS2 & 0.0069   &              &                                  & 0.0041   &                                                     \\ \hline
RLF & 0.01192   &              &                                  & 0.0037   &                                                     \\ \hline
RRAGB & 0.0350   &              &                                  & 0.00978   &                                                     \\ \hline
SCGB2A1 & 0.0133   &              &                                  & 0.0023  &                                                     \\ \hline
PLI & 0.0044   &              &                                  &               &                                                     \\ \hline
SH2D7 & 0.0088   &              &                                  &               &                                                     \\ \hline
SKAP2 & -0.0165  &              &                                  & -0.0048  &                                                     \\ \hline
SLC25A43 & -0.0155 &              &                                  & -0.0021  &                                                     \\ \hline
SLC7A8 & -0.0118  &              &                                  & -0.0029  &                                                     \\ \hline
SMARCD3 & -0.0144  &              &                                  & -0.0074  &                                                     \\ \hline
SPAG16 & -0.0115  &              &                                  & -0.0023  &                                                     \\ \hline
TAAR8 & 0.0036   &              &                                  & 0.0002  &                                                     \\ \hline

\end{tabular}
\end{table}

\begin{table}[]
\caption{Continued analysis of the LUAD data using CER: identified main effects and interactions}
\begin{tabular}{|l|l|l|l|l|l|}
\hline
Gene     &               & Sex          & Patient Smoking History& Diagnosis Age & Cancer.Tumor.Stage \\ \hline

TGM7 & -0.0064  &              &                                  & -0.0014  &                                                     \\ \hline
TMEM182 & 0.0114   &              &                                  & 0.0049   &                                                     \\ \hline
TRIM63 & -0.0128  &              &                                  & -0.0034  &                                                     \\ \hline
TRIT1 & -0.0082  &              &                                  & -0.0046  &                                                     \\ \hline
UNC5D & -0.0002 &              &                                  &               &                                                     \\ \hline
VEGFC & -0.0086  &              &                                  & -0.0062  &                                                     \\ \hline
ZNF641 & -0.0090 &              &                                  & -0.0007 &                                                     \\ \hline
ZRANB2 & 0.0104   &              &                                  & 0.0070   &                                                     \\ \hline
\end{tabular}
\end{table}
\newpage

%
%






\bibliographystyle{plain}

\end{document}